\documentclass[pra,aps,superscriptaddress,twocolumn,floatfix]{revtex4-1}

\usepackage{dsfont}
\usepackage{mathtext}

\usepackage{mathtools}
\usepackage[usenames,dvipsnames]{xcolor}
\usepackage{amsmath}
\usepackage{amssymb,mathrsfs}
\usepackage{graphicx}
\usepackage{epstopdf}

\usepackage{mathtext}

\newcommand{\la}{\lambda}

\newcommand{\om}{\omega}

\newcommand{\prt}{\partial}

\newcommand{\sn}{\mathrm{sn}}
\newcommand{\cn}{\mathrm{cn}}

\graphicspath{{Figures/}}

\begin{document}

\title{Riemann problem for the light pulses in optical fibers for the generalized Chen-Lee-Liu equation}

\author{
Sergey K. Ivanov\\
~\\
\small{\it Moscow Institute of Physics and Technology, Institutsky lane 9, Dolgoprudny, Moscow region, 141700, Russia}\\
\small{\it Institute of Spectroscopy, Russian Academy of Sciences, Troitsk, Moscow, 108840, Russia}
}

\begin{abstract}
We provide the classification of possible wave structures evolving from initially
discontinuous profiles for the photon fluid
propagating in a normal dispersion fiber. The dynamics
of light field is described by the generalized Chen-Lee-Liu equation,
which belongs to the family of the nonlinear Schr\"odinger equations with a self-steepening type term
appearing due to retardation of the fiber material response to variations of the electromagnetic
signal. This equation is also used in investigations of the dynamics of modulated waves propagating through
a single nonlinear transmission network. We describe its periodic solutions and
the corresponding Whitham modulation equations. The wave patterns generated by the initial parameter
profiles are composed of different building blocks which are presented in detail.
It is shown that evolution dynamics in this case is much richer than that for the nonlinear Schr\"odinger
equation. Complete classification of possible wave structures is given for all possible jump conditions
at the discontinuity. Our analytic results are confirmed by numerical simulations.

~

\textbf{Keywords:} nonlinear medium, optical fibers, Chen-Lee-Liu equation, dispersive shock waves, solitons, Whitham modulation equations.
\end{abstract}



\maketitle

\section{Introduction}\label{sec1}

The evolution of light pulses in waveguides is a subject of active modern experimental
and theoretical research. One of the trends in this field is the study of so-called dispersive shock waves. Such structures are observed in various physical media, such as water waves (where such waves are often called undular bores), Bose-Einstein condensates, waves in magnetics, in nonlinear optics, and other areas of physics (see, e.g., \cite{kamch-2000,eh-2016}). It is well known that if one neglects the effects of dissipation and dispersion,
then the theory of nonlinear propagation
of light envelopes suffers from a wave-breaking singularity developed at some
finite fiber length after which
a formal solution of nonlinear wave equations becomes
multi-valued and loses its physical meaning.
The account of dispersion eliminates such a non-physical behavior. But the evolution equations for the envelopes acquire
higher order derivatives, and after the wave-breaking
moment, instead of the multi-valued region, an expanding
region of fast nonlinear oscillations is formed. Envelope
parameters in such a structure change slowly compared
with the characteristic oscillations frequency and their
wavelength. This region of fast oscillations is called
the ``dispersive shock wave'' (DSW).

One of the most substantial problems in which DSWs
can occur is the \textit{Riemann problem}, which includes classification of wave
structures resulting from the evolution of the initial discontinuity.
This problem has played an important role since the classical paper
of B.~Riemann \cite{riem-1860}, subsequently supplemented by the jump conditions of W.~Rankin \cite{rank-1870}
and H.~Hugoniot \cite{hug-1887,hug-1889}; it served as a prototype of the example of
shock formation in dispersionless media with small viscosity. The
full classification of possible wave patterns evolving from initial discontinuities
was obtained by N.~Kotchine \cite{kot-1926}. However, in the case of optical systems,
we have dispersion instead
of viscosity. The problem where DSWs are formed instead of viscous shocks
was studied for
the first time in the context of the physics of shallow water waves
which evolution is described by the celebrated Korteweg-de Vries (KdV)
equation \cite{kv-1895}. The equations governing the slow
evolution of the envelope of the nonlinear oscillations had been
derived by G.~B.~Whitham \cite{whith-1965} and they were applied to the
description of the DSW structure by A.~V.~Gurevich and L.~P.~Pitaevskii
\cite{gp-1973}. Because of the universality of the KdV equation, this
approach can naturally be applied to many other physical situations.
Later it became clear \cite{fmf-1980} that the
diagonalization of the Whitham modulation equations is possible due to the
special property discovered in \cite{ggkm-1967} of the complete integrability of the
KdV equation. Development of the finite-gap integration method
\cite{lax-1974,nov-1974}, as well as the methods of deriving \cite{fmf-1980,kric-1988}
and solving
the Whitham equations \cite{tsar-1991,dn-1993}, made it possible to extend the Gurevich-Pitaevskii
approach to a number of other completely \textit{integrable} equations of physical
interest (see, for instance, \cite{eh-2016}) and it admits investigation of
\textit{non-genuinely nonlinear} hyperbolic systems.

In nonlinear optics, DSWs were observed long ago
(see, e.g., \cite{tsj-1985,rg-1989}), but they remain
the subject of active current experimental (see, e.g., \cite{xmktcc-2016,nfxmef-2019,mpaldc-2019})
and theoretical (see, e.g., \cite{ikp-2019,mpggcc-2019}) research.
In the fiber optics applications, the dynamics of pulses is usually described
by the nonlinear Schr\"odinger (NLS) equation that accounts for two main effects---quadratic normal dispersion
and Kerr nonlinearity. For this case, the theory of DSWs is already well developed and the main parameters
of the arising wave structures can be calculated for typical idealized situations in simple
analytical form. Extension of the Gurevich-Pitaevskii
approach to the NLS equation became possible only after derivation
of the Whitham modulation equations \cite{fl-1986,pav-1987} by the methods based on the inverse
scattering transform for the NLS equation \cite{zs-1971} which means its complete integrability.
In particular, consideration of many realistic problems can be reduced to
analysis of the so-called Riemann problem of evolution of discontinuity in the initial data.
Such a discontinuity can appear, for example, as a jump in the time dependence of the
light intensity, which is mostly typical in physics of light pulses in fibers, or evolve from
a ``collision'' of two pulses in which case not only intensity has a discontinuity but
also the time and space derivatives of the phase. Classification of possible wave structures
in the NLS equation theory was given in Refs.~\cite{gk-1987,el-1995}, and it provides the theoretical
basis for calculation of characteristic parameters of such experiments as that of Ref.~\cite{xckmt-2017}.
It was shown that the NLS theory evolution of any initial discontinuity leads to a wave pattern
consisting of a sequence of building blocks two of which are represented by either the rarefaction wave
or the DSW.
However, in nonlinear optics, besides quadratic dispersion and Kerr nonlinearity, many other
effects can play important role in the propagation of pulses. For instance, in experiment \cite{wjf-2007}
with photorefractive material the saturation of nonlinearity is quite essential and the
corresponding theory of DSWs was developed in Ref.~\cite{el-2007}. Naturally,
in order to get a better
understanding of the higher-order nonlinear effects, it is
necessary to introduce several other higher-order terms, such
as third-order dispersion, quintic nonlinear terms, etc., into the
Hirota equation \cite{Hirota-1973}, Kundu-Eckhaus equation \cite{Kundu-1984,Calogero-1987},
Lakshmanan-Porsezian-Daniel equation \cite{Porsezian-1992},
and a generalized NLS equation \cite{Wang-2013}.
Due to recent technological developments in laser and ultra-high-bit-rate optical fiber
communication, these higher-order nonlinear effects are
unavoidable in many optical systems when modeling the
transmission of ultra-short and high-intensity light pulses in
nonlinear optical media. In fiber optics, one
needs to take into account such effects as dissipation, higher-order dispersion, intra-pulse
Raman scattering and self-steepening (see, e.g., \cite{ka-2003}). These effects can drastically
change evolution of DSWs leading sometimes to violation of the supposition that such an evolution is adiabatically slow.
Theoretically, the self-steepening term that arise in optical setting is commonly associated with the class
of the derivative nonlinear Schr\"odinger (DNLS) equation. It is relevant to mention that the self-steepening
of an optical fiber pulse, otherwise called Kerr dispersion, arises when the group velocity of a pulse
depends on the intensity \cite{Han-2011}. Several versions of the DNLS equations have been studied from different points of
view. Well known NLS equations with derivative terms include the Kaup-Newell equation \cite{Kaup-1978}, the
Chen-Lee-Liu equation \cite{Chen-1979}, and the Gerjikov-Ivanov equation \cite{Gerdjikov-1983} which arise in theories of nonlinear
optics, fluid dynamics and plasma physics. These nonlinear wave equations are usually called
DNLS-{\MakeUppercase{\romannumeral 1}}, DNLS-{\MakeUppercase{\romannumeral 2}} and DNLS-{\MakeUppercase{\romannumeral 3}}
equations, respectively. The pulse propagation in a single-mode optical fiber can be described
by the Chen-Lee-Liu (CLL) equation
\begin{equation} \label{CLL}
    \begin{split}
    i q_t+\frac{1}{2}q_{xx}+ i \delta |q|^2q_x=0,
    \end{split}
\end{equation}
where the coordinates $t$ and $x$ denote propagation distance and retarded time,
but represent slow time and spatial coordinate traveling with group velocity
in hydrodynamics, respectively. In optical fiber setting, the term involving parameter $\delta>0$
is usually associated with the self-steepening phenomena \cite{Rogers-2012}. In Ref.~\cite{Moses-2007}
J.~Moses et al. performed optical pulse propagation involving self-steepening without self-phase-modulation.
This experiment provides the first experimental evidence of the CLL equation.
The CLL equation (\ref{CLL}) corresponds to a situation where the dispersionless
Riemann invariants depend nonmonotonically on the
physical variables; that is, the problem is {not-genuinely
nonlinear} (see, e.g., \cite{lax-2006}). In this case, new types of wave structures arise,
which are similar to contact discontinuities in the theory of viscous shock waves.
So the CLL equation can be considered as an example of \textit{non-convex dispersive hydrodynamics}.

We consider the propagation of an optical pulse inside a monomode fiber modeled by
a \textit{generalized Chen-Lee-Liu} (generalized CLL) equation,
\begin{equation} \label{gCLL}
\begin{split}
    i q_t+\frac{1}{2}q_{xx} -|q|^2q + i \delta |q|^2q_x=0,
\end{split}
\end{equation}
that also includes the Kerr nonlinearity, which is inevitable at sufficiently high intensities.
The generalized CLL equation is simply
related to the CLL equation (see, e.g., \cite{Forest-2009}). This equation is also used in the
investigation of modulated wave dynamics of waves propagating through a single nonlinear
transmission network, which also presents practical interest (see \cite{Lin-2019} and references therein).

Motivated by applications of the generalized CLL equation (\ref{gCLL}),
we consider the method which permits one to predict a wave pattern
arising from any given data for an initial discontinuity. The method is quite general and
it was applied to the generalized NLS equations with self-steepening nonlinearity \cite{ik-2017,kamch-2018}
and to the Landau-Lifshitz equation for magnetics with easy-plane anisotropy (or polarization waves in a two-component
Bose-Einstein condensate) \cite{ikcp-2017}. Here we extend the
theory to the not-genuinely case of the generalized CLL equation (\ref{gCLL}).
The exact integrability of this equation makes it possible to develop
a Whitham modulational theory for describing
configurations where nonlinear waves are slowly modulated, as observed in dispersive shocks.

The paper is organized as follows: The influence of the last term of the
generalized CLL equation (\ref{gCLL}) on
the dynamics of linear waves that propagate along a
uniform background and small-amplitude limit
is discussed in Sec.~\ref{sec2}.
The exact integrability of this
equation is used in Sec.~\ref{sec3} for derivation of the Whitham modulational equations.
In Sec.~\ref{sec4} we describe the elementary wave structures that
appear as building blocks in the general wave patterns.
The full classification of the solutions of the Riemann problem is presented in Sec.~\ref{sec5}.
The last Sec.~\ref{sec6} is devoted to conclusions.

\section{Linear waves and small dispersion and weak nonlinearity limits}\label{sec2}

Let us turn to the study of linear waves
and
the small-amplitude and weak-dispersion limits
when these two effects are taken into account in the main approximation.
That is, we are interested in propagation of disturbances along a uniform background intensity $ \rho_0 $.
To this end,
it is convenient to use the physical variables of intensity
$ \rho (x, t) $ and chirp $ u (x, t)$.
To go to the equations for these variables, we apply the Madelung transform
\begin{equation} \label{Madelung}
    q(x,t) = \sqrt{\rho(x,t)}\,\exp{\left( i \int^x u(x',t)dx'\right)}.
\end{equation}
After its substitution into Eq.~(\ref{gCLL}),
separation of the real and imaginary parts
and differentiation of one of the equations with respect to $x$,
we get the system
\begin{equation} \label{Hydro}
\begin{split}
    & \rho_t+(\rho u +\frac12 \delta \rho^2)_x=0, \\
    & u_t+uu_x+\rho_x+\delta(\rho u)_x+\left(\frac{\rho^2_x}{8\rho^2}-\frac{\rho_{xx}}{4\rho}\right)_x=0.
\end{split}
\end{equation}
The last term on the left-hand side of the second equation
describes the dispersion.

The linear dispersion relation of the system (\ref{Hydro}) for linear waves propagating along a constant background $\rho_0$ has the form
\begin{equation} \label{DispersionLaw}
    \omega_{1,2}(k)= \left(\delta\rho_0 \pm  \sqrt{\frac{k^2}{4}+\rho_0}\right)k.
\end{equation}
Here $\omega$ is the frequency of the linear waves and $k$ is the
wave number.
Suppose that at the initial moment the phase of the wave
is constant and there is only the intensity perturbation.
After standard calculations we get the solution of the linear problem expressed in terms of the
Fourier transform $\widehat{\rho_0}'(k)$ of the initial (input) intensity linear disturbance,
\begin{equation} \label{Intensity_perturbation}
\begin{split}
    \rho'(x,t)=\frac{1}{4\pi}\int_{-\infty}^{+\infty} \widehat{\rho_0}'(k)
    \left[e^{i(kx-\omega_1t)}+e^{i(kx-\omega_2t)} \right] dk.
\end{split}
\end{equation}
One can see that an initial pulse splits into two smaller pulses; however,
in contrast to the NLS case, two pulses propagate
with different group velocities. This is a manifestation of
lack of the $x$-inversion invariance, which is caused by the last term in the generalized CLL
equation (\ref{gCLL}).

We are interested in
the leading dispersion and nonlinear corrections to the
dispersionless linear propagation of disturbances.
Using the system (\ref{Hydro}) and applying the standard perturbation
theory for the amplitude of the perturbation
and for the weak dispersion (see, for example, \cite{kamch-2000}),
one can obtain a small-amplitude
analog of Eq.~(\ref{gCLL}). Let the wave propagate
in the positive
direction of the $x$ axis. Then an approximate equation for
$ \rho '= \rho-\rho_0 $ takes the form
\begin{equation}\label{KdVLimit}
\begin{split}
    \frac{\prt \rho'}{\prt t}&+\left(\delta\rho_0+\sqrt{\rho_0}\right)\frac{\prt \rho'}{\prt x}\\
    &+\frac32\frac{1+\delta\sqrt{\rho_0}}{\delta\sqrt{\rho_0}}\delta\rho'\frac{\prt \rho'}{\prt x}
    -\frac{1}{8\sqrt{\rho_0}}\frac{\prt^3 \rho'}{\prt x^3}=0.
\end{split}
\end{equation}
This is the Korteweg-de Vries (KdV) equation.
Formation of DSWs from initial discontinuities in the KdV equation theory has been well known since
the pioneering paper Ref.~\cite{gp-1973}: the initial discontinuity evolves into either
a rarefaction wave or a cnoidal DSW.
In the limit $\delta\sqrt{\rho}\rightarrow1$ the nonlinear term of
Eq.~(\ref{KdVLimit}) has finite value and
we need not include higher-order corrections for taking
into account higher-order nonlinear effects.
The situation is the opposite for another simple wave
propagating in the negative
direction of the $x$ axis.
In this case we have the Gardner equation
\begin{equation}\label{GardnerLimit}
\begin{split}
    \frac{\prt \rho'}{\prt t}&+\left(\delta\rho_0-\sqrt{\rho_0}\right)\frac{\prt \rho'}{\prt x}
    -\frac32\frac{1-\delta\sqrt{\rho_0}}{\delta\sqrt{\rho_0}}\delta\rho'\frac{\prt \rho'}{\prt x}\\
    &+\frac38\frac{1+\delta\sqrt{\rho_0}}{\rho_0^{3/2}}\rho'^2\frac{\prt \rho'}{\prt x}
    +\frac{1}{8\sqrt{\rho_0}}\frac{\prt^3 \rho'}{\prt x^3}=0.
\end{split}
\end{equation}
In the limit $\delta\sqrt{\rho}\rightarrow1$ the last equation reduces to the
modified Korteweg-de Vries (mKdV) equation
\begin{equation}\label{mKdVLimit}
\begin{split}
    \frac{\prt \rho'}{\prt t}+\frac{3}{4\rho_0^{3/2}}\rho'^2\frac{\prt \rho'}{\prt x}
    +\frac{1}{8\sqrt{\rho_0}}\frac{\prt^3 \rho'}{\prt x^3}=0.
\end{split}
\end{equation}
The situation for the mKdV and Gardner equations is much more
complicated than for the KdV case \cite{kamch-2012} and in this case we can get eight different structures
including, besides the rarefaction waves and cnoidal DSWs, also trigonometric DSWs,
combined shocks and their combinations separated by plateaus. Therefore one should expect
that in the case of the Riemann problem for Eq.~(\ref{gCLL}) we also have to get much richer
structure than in the NLS case. To solve this problem, at first we have to find periodic
solutions of Eq.~(\ref{gCLL}) in a form that is convenient for us, that is, in a form parametrized by the
parameters related to the Riemann invariants of the corresponding Whitham modulation
equations by simple formulas. In the next section we shall obtain the
periodic solutions by this method and derive the Whitham equations.

\section{Periodic solutions and Whitham modulation equations}\label{sec3}

The finite-gap integration method (see, e.g., \cite{kamch-2000}) is based on the possibility
of representing the generalized CLL equation (\ref{gCLL}) as a compatibility condition of two
systems of linear equations with a spectral parameter $\la$,
\begin{align}
            \label{lax1}
    &\frac{\partial}{\partial x}
    \begin{pmatrix}
        {\psi}_1 \\
        {\psi}_2 \\
    \end{pmatrix}
    =\left(
    \begin{array}{cc}
        {F} & {G} \\
        {H} & -{F} \\
    \end{array}
    \right)
    \begin{pmatrix}
        {\psi}_1 \\
        {\psi}_2 \\
    \end{pmatrix}\;, \\
            \label{lax2}
    &\frac{\partial}{\partial t}
    \begin{pmatrix}
        {\psi}_1 \\
        {\psi}_2 \\
    \end{pmatrix}
    =\left(
    \begin{array}{cc}
        {A} & {B} \\
        {C} & -{A} \\
    \end{array}
    \right)
    \begin{pmatrix}
        {\psi}_1 \\
        {\psi}_2 \\
        \end{pmatrix}\;,
\end{align}
where
\begin{equation}\label{}
    \begin{split}
    & {F}=-\frac{i}{2\delta}\left(\lambda^2-\delta^2|q|^2+1\right),\quad {G}=-q\lambda,\quad {H}=q^*\lambda, \\
    & {A}=-\frac{i}{4\delta^2}\left(\lambda^2-\delta^2|q|^2+1\right)^2-\frac{\delta}{4}\left(q_xq^*-qq_x^*\right)-\frac{i}{2}|q|^2, \\
    & {B}=-\frac{1}{2\delta}\left(\lambda^2-\delta^2|q|^2+1\right)q\lambda-\frac{i}{2}q_x\lambda, \\
    & {C}=\frac{1}{2\delta}\left(\lambda^2-\delta^2|q|^2+1\right)q^*\lambda-\frac{i}{2}q_x^*\lambda.
    \end{split}
\end{equation}
This Lax pair can be obtained by simple transformation from the known Lax
pair for the CLL equation (\ref{CLL}) (see Ref.~\cite{WadatiSogo-1983}).
The $2\times2$ linear problems \eqref{lax1} and
\eqref{lax2} have two linearly independent
basis solutions which we denote as $(\psi_1,\,\psi_2)^T$ and
$(\varphi_1,\,\varphi_2)^T$. We define the ``squared basis functions''
\begin{equation}\label{}
    \begin{split}
    {f}=-\frac{ i}2(\psi_1\varphi_2+\psi_2\varphi_1), \quad
    {g}=\psi_1\varphi_1,          \quad
    {h}=-\psi_2\varphi_2,
    \end{split}
\end{equation}
which obey the linear equations
\begin{subequations}\label{fght}
    \begin{align}
    {f}_x=& { i}G h-{ i}H g, \label{ft} \\
    {g}_x=& 2F g+2{ i}G f, \label{gt} \\
    {h}_x=& -2F h-2{ i}H f, \label{ht}
    \end{align}
\end{subequations}
and
\begin{subequations}\label{fghx}
    \begin{align}
    {f}_t=& { i}B h-{ i}C g, \label{fx} \\
    {g}_t=& 2A g+2{ i}B f, \label{gx} \\
    {h}_t=& -2A h-2{ i}C f. \label{hx}
    \end{align}
\end{subequations}
We look for the solutions of these equations in the form
\begin{equation}\label{sol:fgh}
    \begin{split}
    {f} & = \left(\lambda^2-\delta^2|q|^2+1\right)^2 - {f}_1 \left(\lambda^2-\delta^2|q|^2+1\right) + {f}_2, \\
    {g} & = -2\delta \left(\lambda^2-\delta^2|q|^2+1-\mu\right) q\lambda, \\
    {h} & = 2\delta \left(\lambda^2-\delta^2|q|^2+1-\mu^*\right) q^*\lambda.
    \end{split}
\end{equation}
Here the functions ${f}_1(x,t)$, ${f}_2(x,t)$, $\mu(x,t)$, and $\mu^*(x,t)$
are unknown;  $\mu(x,t)$ and $\mu^*(x,t)$ are not interrelated {\it a priori}, but we
shall find soon that they are complex conjugate, whence the notation.

Substitution of Eqs.~(\ref{sol:fgh}) into Eqs.~(\ref{fght})
gives after equating the coefficients of like powers of $\lambda$
expressions for the $x$-derivatives of ${f}_1$ and ${f}_2$,
\begin{equation}\label{f1t:f2t}
    \begin{split}
    f_{1,x} = 0, \qquad f_{2,x}=\delta^2\left(2-2\delta^2|q|^2-f_1\right)\left(|q|^2\right)_x,
    \end{split}
\end{equation}
and of $|q|^2$ and $\mu$,
\begin{equation}\label{f1t:f2t2}
    \begin{split}
    \left(|q|^2\right)_x & =\frac{i}{\delta}\left(\mu-\mu^*\right)|q|^2, \\
    (\mu q)_x & = - i\delta\left(\mu-\mu^*\right)|q|^2-\frac{i}{\delta}f_2q.
    \end{split}
\end{equation}
In a similar way, substitution of (\ref{sol:fgh}) into (\ref{fghx}) with account of (\ref{f1t:f2t})
gives equations for the $t$-derivatives of ${f}_1$ and ${f}_2$,
\begin{equation}\label{eq42n}
    \begin{split}
    f_{1,t} = 0, \qquad f_{2,t}=\frac{f_1}{2\delta}f_{2,x},
    \end{split}
\end{equation}
and
\begin{equation}\label{eqintder}
    \begin{split}
    \left(|q|^2\right)_t = \frac{f_1}{2\delta} \left(|q|^2\right)_x.
    \end{split}
\end{equation}

It is easy to check that the expression $f^2-gh=P(\lambda)$ does not depend on
$x$ and $t$; however it can depend on the spectral parameter $\la$.
We are interested in the one-phase periodic solution.
It is distinguished by the condition that $P(\lambda)$
is an eighth-degree polynomial of the
form
\begin{equation}\label{}
    \begin{split}
    {f}^2-{g}{h} & = P(\lambda) = \prod_{i=1}^4\left(\lambda^2-\lambda_i^2\right) \\
    & = \lambda^8-s_1 \lambda^6+s_2 \lambda^4-s_3\lambda^2+s_4.
    \end{split}
\end{equation}
Equating the coefficients of like powers of
$\lambda$ at two sides of this identity, we get
\begin{subequations}\label{s1234}
    \begin{align}
    s_1 & = 2f_1-4 \label{s1} , \\
    \begin{split} \label{s2} 
    s_2 & = {f}_1^2+2{f}_2-4\delta^2|q|^2\left(\mu+\mu^*\right) \\ & \quad + 2\left[3f_1-3-\delta^2|q|^2\right]\left[\delta^2|q|^2-1\right],
    \end{split}
     \\
    \begin{split}
    s_3 & = \left(s_2+f_1^2+2f_2+2\left[\delta^4|q|^4-1\right]\right)\left[\delta^2|q|^2-1\right] \\
    & \quad +2f_1f_2 -4\delta^2|q|^2 \mu\mu^*,
    \end{split} \label{s3}
     \\
    s_4 & = \left(\left[f_1+\delta^2|q|^2-1\right]\left[\delta^2|q|^2-1\right]+f_2\right)^2.
    \label{s4}
    \end{align}
\end{subequations}
Here $s_i$ are standard symmetric functions of the four
zeros $\lambda_i^2$ of the polynomial $P(\lambda)$:
\begin{equation}\label{}
    \begin{split}
    & s_1=\sum_i\lambda_i^2,\quad s_2=\sum_{i<j}\lambda_i^2\lambda_j^2,
    \quad s_3=\sum_{i<j<k}\lambda_i^2\lambda_j^2\lambda_k^2, \\
    & s_4=\lambda_1^2\lambda_2^2\lambda_3^2\lambda_4^2.
    \end{split}
\end{equation}

Eqs.~(\ref{s1234}) allow us to express $\mu,\mu^*$ as functions of $|q|^2$.
The first and the last Eqs.~(\ref{s1234}) give
\begin{equation}\label{f1f2}
    \begin{split}
    {f}_1 & = \frac{s_1}{2}+2, \\
    {f}_2 & = 1-\delta^4|q|^4-\frac{s_1}{2}\left(\delta^2|q|^2-1\right)\pm\sqrt{s_4}.
    \end{split}
\end{equation}
We substitute that into (\ref{s2}) and (\ref{s3}) and obtain
the system for $\mu$ and $\mu^*$, which can be easily solved to give
\begin{equation}\label{muR}
    \begin{split}
    \mu =& \frac{1}{2\delta^2|q|^2}\Big[\left(\frac{s_1}{4}\right)^2 + \frac12\delta^2|q|^2\left(s_1+4-2\delta^2|q|^2\right) \\ & - \frac{s_2}{4} \pm\frac12\sqrt{s_4}
    -{ i}\sqrt{-\mathcal{R}\left(\delta^2 |q|^2\right)}\Big],
    \end{split}
\end{equation}
where
\begin{equation}\label{tildeR}
    \begin{split}
    \mathcal{R}(\nu)= & \nu^4+s_1\nu^3+\left(\frac38s_1^2-\frac{s_2}{2}\mp3\sqrt{s_4}\right)\nu^2 \\
                & + \left(\frac{1}{16}s_1^3+s_3-\frac14s_1s_2\mp\frac12s_1\sqrt{s_4}\right)\nu\\
                &+\left(\left(\frac{s_1}{4}\right)^2 - \frac{s_2}{4}\pm\frac12\sqrt{s_4}\right)^2.
    \end{split}
\end{equation}
The function $\mathcal{R}$ introduced here is a fourth-degree polynomial in $\nu$
and it is called an {\it algebraic resolvent} of the polynomial $P(\lambda)$,
because zeros of $\mathcal{R}(\nu)$  are
related to zeros of $P(\lambda)$ by the following simple symmetric
expressions: the upper sign in (\ref{tildeR}) corresponds to the
zeros
\begin{equation}\label{zeros1}
    \begin{split}
    \nu_1 & = \frac14(-\lambda_1+\lambda_2+\lambda_3-\lambda_4)^2, \\
    \nu_2 & = \frac14(\lambda_1-\lambda_2+\lambda_3-\lambda_4)^2, \\
    \nu_3 & = \frac14(\lambda_1+\lambda_2-\lambda_3-\lambda_4)^2, \\
    \nu_4 & = \frac14(\lambda_1+\lambda_2+\lambda_3+\lambda_4)^2. \\
    \end{split}
\end{equation}
and the lower sign in Eq.~(\ref{tildeR}) corresponds to the zeros
\begin{equation}\label{zeros2}
    \begin{split}
    \nu_1 & = \frac14(-\lambda_1+\lambda_2+\lambda_3+\lambda_4)^2, \\
    \nu_2 & = \frac14(\lambda_1-\lambda_2+\lambda_3+\lambda_4)^2, \\
    \nu_3 & = \frac14(\lambda_1+\lambda_2-\lambda_3+\lambda_4)^2, \\
    \nu_4 & = \frac14(\lambda_1+\lambda_2+\lambda_3-\lambda_4)^2,
    \end{split}
\end{equation}
This can be proved by a simple check of the Vi\`ete formulas.

As follows from the second Eqs.~(\ref{eq42n}), the first Eq.~(\ref{f1f2}) gives the expression for the constant phase velocity,
\begin{equation}\label{PhaseVeloc}
    \begin{split}
    V & =-\frac{f_1}{2\delta}=-\frac{1}{\delta} - \frac{s_1}{4\delta} \\ & =-\frac{1}{\delta}-\frac{1}{4\delta}\sum_{i=1}^4\la^2_i=-\frac{1}{\delta}-\frac{1}{4\delta}\sum_{i=1}^4\nu_i.,
    \end{split}
\end{equation}
and we find that $f_2$ depends on $\xi=x-Vt$ only. Then from Eq.~(\ref{eqintder}) we see that
the intensity $\rho=|q|^2$ also depends only on $\xi$. The equations for dynamics of $\rho$ can be easily
found by substitution of (\ref{muR}) into the first Eq.~(\ref{f1t:f2t2}), so we get
\begin{equation}\label{}
    \begin{split}
    \frac{d(\delta^2\rho)}{d\xi}=\frac{1}{\delta}\sqrt{-\mathcal{R}(\delta^2\rho)},
    \end{split}
\end{equation}
where $\mathcal{R}$ is, as we know, a fourth degree polynomial with the zeros given in terms of $\la_i$
by the formulas (\ref{zeros1}) or (\ref{zeros2}). This equation can be solved in a standard way in terms of
elliptic functions. Without going to much detail we shall present here the main results.

We shall assume that $\la_i$ are ordered according to $0\le\la_1\leq \la_2\leq\la_3\leq\la_4$ and then both our definitions (\ref{zeros1}) and (\ref{zeros2}) give the same ordering
of $\nu_i$: $\nu_1\leq\nu_2\leq\nu_3\leq\nu_4$. The real solutions correspond to oscillations of $\delta^2 \rho$ within the intervals where
$-\mathcal{R}(\delta^2 \rho)\geq0$.

(A) At first we shall consider the periodic solution corresponding to oscillations
of $\delta^2 \rho$ in the interval
\begin{equation}\label{eq18}
    \nu_1\leq \delta^2 \rho\leq \nu_2.
\end{equation}
Standard calculation yields, after some algebra, the solution in terms of Jacobi
elliptic functions:
\begin{equation}\label{eq20}
    \delta^2 \rho=\nu_2-\frac{(\nu_2-\nu_1)\cn^2(\theta,m)}{1+\frac{\nu_2-\nu_1}{\nu_4-\nu_2}\sn^2(\theta,m)},
\end{equation}
where it is assumed that $\delta^2 \rho(0)=\nu_1$,
\begin{equation}\label{eq21}
    \theta=\sqrt{(\nu_3-\nu_1)(\nu_4-\nu_2)}\,\xi/(2\delta),
\end{equation}
\begin{equation}\label{eq22}
    m=\frac{(\nu_4-\nu_3)(\nu_2-\nu_1)}{(\nu_4-\nu_2)(\nu_3-\nu_1)},
\end{equation}
the functions $\cn$ and $\sn$ being Jacobi elliptic functions \cite{AbramowitzStegun-72}.
The period of oscillations along the $x$ axis is equal to function \eqref{eq20} is
\begin{equation}\label{eq23}
    L=\frac{4K(m)}{\sqrt{(\nu_3-\nu_1)(\nu_4-\nu_2)}}=\frac{4K(m)}{\sqrt{(\la_3^2-\la_1^2)(\la_4^2-\la_2^2)}},
\end{equation}
where $K(m)$ is the complete elliptic integral of the first kind
\cite{AbramowitzStegun-72}.

In the limit $\nu_3\to \nu_2$ ($m\to1$) the period
tends to infinity and the solution (\ref{eq20}) acquires the soliton form
\begin{equation}\label{eq24}
    \delta^2 \rho=\nu_2-\frac{\nu_2-\nu_1}{\cosh^2\theta+\frac{\nu_2-\nu_1}{\nu_4-\nu_2}\sinh^2\theta}.
\end{equation}
This is a ``dark soliton'' for the variable $\rho$.

The limit $m\to0$ can be reached in two ways.

(i) If $\nu_2\to \nu_1$, then the solution transforms into a linear harmonic
wave,
\begin{equation}\label{eq25}
    \begin{split}
    \delta^2 \rho&\cong \nu_2-\frac12(\nu_2-\nu_1)\cos(\om\xi/\delta),\\
    \om&=\sqrt{(\nu_3-\nu_1)(\nu_4-\nu_1)}.
\end{split}
\end{equation}

(ii) If $\nu_4=\nu_3$ but $\nu_1\neq \nu_2$, then we arrive at
the nonlinear trigonometric solution:
\begin{equation}\label{eq26}
    \begin{split}
    \delta^2 \rho&=\nu_2-\frac{(\nu_2-\nu_1)\cos^2\theta}{1+\frac{\nu_2-\nu_1}{\nu_3-\nu_2}\sin^2\theta},\\
    \theta&=\sqrt{(\nu_3-\nu_1)(\nu_3-\nu_2)}\,\xi/(2\delta).
    \end{split}
\end{equation}
If we take the limit $\nu_2-\nu_1\ll \nu_3-\nu_1$ in this solution, then we
return to the small-amplitude limit (\ref{eq25}) with $\nu_4=\nu_3$. On
the other hand, if we take here the limit $\nu_2\to \nu_3=\nu_4$, then the
argument of the trigonometric functions becomes small and we can
approximate them by the first terms of their series expansions. This
corresponds to an algebraic soliton of the form
\begin{equation}\label{eq27}
   \delta^2 \rho=\nu_2-\frac{\nu_2-\nu_1}{1+(\nu_2-\nu_1)^2\xi^2/(4\delta^2)}.
\end{equation}

(B) In the second case, the variable $\delta^2 \rho$ oscillates in the interval
\begin{equation}\label{eq28}
    \nu_3\leq \delta^2 \rho\leq \nu_4\; .
\end{equation}
Here again, a standard calculation yields
\begin{equation}\label{eq30}
   \delta^2 \rho=\nu_3+\frac{(\nu_4-\nu_3)\cn^2(\theta,m)}{1+\frac{\nu_4-\nu_3}{\nu_3-\nu_1}\sn^2(\theta,m)}
\end{equation}
with the same definitions (\ref{eq21}), (\ref{eq22}), and (\ref{eq23})
for $\theta$, $m$, and $T$, correspondingly. In this case we have $\delta^2 \rho(0)=\nu_4$.
In the soliton limit $\nu_3\to \nu_2$ ($m\to1$) we get
\begin{equation}\label{eq31}
   \delta^2 \rho=\nu_2+\frac{\nu_4-\nu_2}{\cosh^2\theta+\frac{\nu_4-\nu_2}{\nu_2-\nu_1}\sinh^2\theta}.
\end{equation}
This is a ``bright soliton'' for the variable $\rho$.

Again, the limit $m\to0$ can be reached in two ways.

(i) If $\nu_4\to \nu_3$, then we obtain a small-amplitude harmonic wave
\begin{equation}\label{eq32}
    \begin{split}
    \delta^2 \rho&\cong \nu_3+\frac12(\nu_4-\nu_3)\cos(\om\xi/\delta),\\
    \om&=\sqrt{(\nu_3-\nu_1)(\nu_3-\nu_2)}.
    \end{split}
\end{equation}

(ii) If $\nu_2=\nu_1$, then we obtain another nonlinear trigonometric solution,
\begin{equation}\label{eq33}
    \begin{split}
    \delta^2 \rho&=\nu_3+\frac{(\nu_4-\nu_3)\cos^2\theta}{1+\frac{\nu_4-\nu_3}{\nu_3-\nu_1}\sin^2\theta},\\
    \theta&=\sqrt{(\nu_3-\nu_1)(\nu_4-\nu_1)}\,\xi/(2\delta).
    \end{split}
\end{equation}
If we assume that $\nu_4-\nu_3\ll \nu_4-\nu_1$, then this reproduces
the small-amplitude limit (\ref{eq32}) with $\nu_2=\nu_1$. On the other hand,
in the limit $\nu_3\to \nu_2=\nu_1$ we obtain the algebraic soliton solution:
\begin{equation}\label{eq34}
    \delta^2 \rho=\nu_1+\frac{\nu_4-\nu_1}{1+(\nu_4-\nu_1)^2\xi^2/(4\delta^2)}.
\end{equation}

The convenience of this form of periodic solutions of our equation is related to the fact that
the parameters $\lambda_i$, connected with $\nu_i$ by the formulas (\ref{zeros1}), (\ref{zeros2}),
play the role of Riemann invariants in the Whitham theory of modulations. For both cases (\ref{zeros1}), (\ref{zeros2})
we have the identities
\begin{equation}\label{}
    m=\frac{(\nu_4-\nu_3)(\nu_2-\nu_1)}{(\nu_4-\nu_2)(\nu_3-\nu_1)}
        =\frac{(\lambda_4^2-\lambda_3^2)(\lambda_2^2-\lambda_1^2)}{(\lambda_4^2-\lambda_2^2)(\lambda_3^2-\lambda_1^2)}.
\end{equation}

Now we shall consider slowly modulated waves. In
this case, the parameters $\lambda_i$ ($i = 1, 2, 3, 4$) become slowly
varying functions of $x$ and $t$ changing little in one period and
they can serve as Riemann invariants.
Evolution of $\la_i$ is governed by the Whitham modulation equations
\begin{equation}\label{WhithamEq}
    \frac{\partial \lambda_i}{\partial t}+v_i\frac{\partial \lambda_i}{\partial x}=0, \quad i = 1,2,3,4.
\end{equation}
The Whitham velocities appearing in these equations can
be computed by means of the formula
\begin{equation}\label{}
    \begin{split}
    {v_i}=\left(1-\frac{L}{\partial_i L}\partial_i\right){V},
    \quad \mbox{where} \quad \partial_i\equiv\frac{\partial}{\partial \lambda^2_i},
    \end{split}
\end{equation}
with the use of Eqs.~(\ref{PhaseVeloc}), (\ref{eq23}). Hence, a simple calculation yields
the explicit expressions
\small
\begin{equation}\label{}
    \begin{split}
    & {v_1}=-\frac{1}{\delta}+\frac{1}{4\delta}\sum_{i=1}^{4}\lambda_i^2
        -\frac{1}{2\delta}\frac{(\lambda_4^2-\lambda_1^2)(\lambda_2^2-\lambda_1^2)K(m)}
        {(\lambda_4^2-\lambda_1^2)K(m)-(\lambda_4^2-\lambda_2^2)E(m)}, \\
    & {v_2}=-\frac{1}{\delta}+\frac{1}{4\delta}\sum_{i=1}^{4}\lambda_i^2
        +\frac{1}{2\delta}\frac{(\lambda_3^2-\lambda_2^2)(\lambda_2^2-\lambda_1^2)K(m)}
        {(\lambda_3^2-\lambda_2^2)K(m)-(\lambda_3^2-\lambda_1^2)E(m)}, \\
    & {v_3}=-\frac{1}{\delta}+\frac{1}{4\delta}\sum_{i=1}^{4}\lambda_i^2
        -\frac{1}{2\delta}\frac{(\lambda_4^2-\lambda_3^2)(\lambda_3^2-\lambda_2^2)K(m)}
        {(\lambda_3^2-\lambda_2^2)K(m)-(\lambda_4^2-\lambda_2^2)E(m)}, \\
    & {v_4}=-\frac{1}{\delta}+\frac{1}{4\delta}\sum_{i=1}^{4}\lambda_i^2
        +\frac{1}{2\delta}\frac{(\lambda_4^2-\lambda_3^2)(\lambda_4^2-\lambda_1^2)K(m)}
        {(\lambda_4^2-\lambda_1^2)K(m)-(\lambda_3^2-\lambda_1^2)E(m)},
    \end{split}
\end{equation}
\normalsize
where $E(m)$ is the complete elliptic integral of the second kind
\cite{AbramowitzStegun-72}.

In a modulated wave representing a dispersive shock wave, the Riemann invariants change
with $x$ and $t$. The dispersive shock wave occupies a space interval at which edges
two of the Riemann invariants coincide.
The soliton edge corresponds to $\la_3=\la_2$ $(m=1)$ and at this edge
the Whitham velocities are given by
\begin{equation}\label{sol-limit}
     \begin{split}
    {v_1} & =-\frac{1}{\delta}+\frac{1}{4\delta}(3\lambda_1^2+\lambda_4^2), \\
    {v_2} & ={v_3}=-\frac{1}{\delta}+\frac{1}{4\delta}(\lambda_1^2+2\lambda_2^2+\lambda_4^2), \\
    {v_4} & =-\frac{1}{\delta}+\frac{1}{4\delta}(\lambda_1^2+3\lambda_4^2).
    \end{split}
\end{equation}
The small amplitude limit $m=0$ can be obtained in two ways. If
$\la_3=\la_4$, then we get
\begin{equation}\label{small-limit}
\begin{split}
    {v_1} & =-\frac{1}{\delta}+\frac{1}{4\delta}(3\lambda_1^2+\lambda_2^2), \quad {v_2}=-\frac{1}{\delta}+\frac{1}{4\delta}(\lambda_1^2+3\lambda_2^2), \\
    {v_3} & ={v_4}=-\frac{1}{\delta}+\frac{1}{\delta}\lambda_4^2+ \frac{1}{4\delta}\frac{(\lambda_2^2-\lambda_1^2)^2}{\lambda_1^2+\lambda_2^2-2\lambda_4^2},
    \end{split}
\end{equation}
and if $\la_2=\la_1$, then
\begin{equation}\label{small-limit2}
\begin{split}
    {v_1} & ={v_2}=-\frac{1}{\delta}+\frac{1}{\delta}\lambda_1^2- \frac{1}{4\delta}\frac{(\lambda_4^2-\lambda_3^2)^2}{\lambda_3^2+\lambda_4^2-2\lambda_1^2}, \\
    {v_3} & =-\frac{1}{\delta}+\frac{1}{4\delta}(3\lambda_3^2+\lambda_4^2), \quad {v_4}=-\frac{1}{\delta}+\frac{1}{4\delta}(\lambda_3^2+3\lambda_4^2).
    \end{split}
\end{equation}

We can now proceed to the description of key elements
(``building blocks'') from which the wave patterns are
constructed.

\section{Key elements}\label{sec4}

We consider in the present paper the so-called Riemann
problem. This corresponds to the study of the evolution
of initial discontinuous profiles of the form
\begin{equation}\label{init}
    \begin{split}
    \rho(t=0)=
    \begin{cases}
        \rho^L, & \quad x<0, \\
        \rho^R, & \quad x>0,
    \end{cases}\quad \\
        u(t=0)=
    \begin{cases}
        u^L, & \quad x<0, \\
        u^R, & \quad x>0.
    \end{cases}\quad
    \end{split}
\end{equation}
Evolution of such a pulse leads to formation of quite complex structures
consisting of simpler elements. We shall describe these elements in the
present section.

\subsection{Rarefaction waves}

For smooth enough wave patterns we can neglect the last dispersion term
in the second equation of the system (\ref{Hydro}) and arrive at
the so-called dispersionless equations
\begin{equation} \label{Hydro2}
\begin{split}
    & \rho_t+(\rho u +\frac12 \delta \rho^2)_x=0, \\
    & u_t+uu_x+\rho_x+\delta(\rho u)_x=0.
\end{split}
\end{equation}
First of all, this system admits a trivial solution for which
$\rho=\mathrm{const}$ and $u=\mathrm{const}$. We shall call such a solution
a ``plateau''. It is
convenient to transform the system (\ref{Hydro2}) to a diagonal Riemann form
\begin{equation}\label{RiemannEquat}
    \begin{split}
    \frac{\partial r_{\pm}}{\partial t}+v_{\pm}\frac{\partial r_{\pm}}{\partial x}=0,
    \end{split}
\end{equation}
by defining the Riemann invariants and Riemann velocities
\begin{equation}\label{RiemannInv}
    \begin{split}
    r_{\pm} = & \frac{u}{2}+\frac{\delta\rho}{2}\pm\sqrt{(1+\delta u)\rho}, \\
    v_{\pm} = & {u}+{\delta\rho}\pm\sqrt{(1+\delta u)\rho},
    \end{split}
\end{equation}
where the Riemann velocities are expressed via the Riemann invariants by the relations
\begin{equation}\label{riemann-vel}
    \begin{split}
    v_+=\frac32r_{+}+\frac12r_{-},\quad  v_-=\frac12r_{+}+\frac32r_{-}.
    \end{split}
\end{equation}
It is clear that the system is modulationally unstable if
\begin{equation}\label{ModulUnstable}
    \begin{split}
    u<-\frac{1}{\delta}.
    \end{split}
\end{equation}

\begin{figure}[t] \centering
\includegraphics[width=0.45\textwidth]{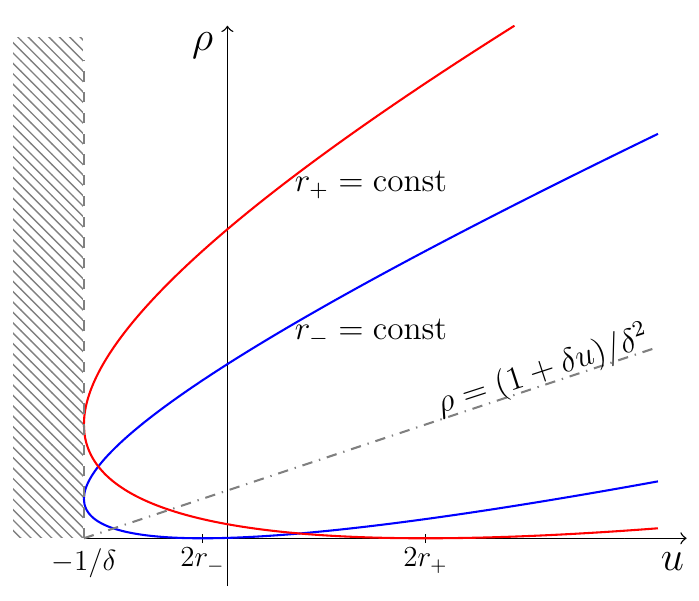}
\caption{Relation between $\rho$ and $u$ for the simple wave solutions in the dispersionless regime.
One line corresponds to $r_-=\mathrm{const}$ (blue), and another one to $r_+=\mathrm{const}$ (red).
Gray area shows the modulationally unstable region $u<-1/\delta$.}
\label{Fig1}
\end{figure}

A rarefaction wave belongs to the class of simple wave solutions.
For such a solution, one of the Riemann invariants is constant,
and this condition either $r_+=\mathrm{const}$ or $r_-=\mathrm{const}$ gives,
when applied to Eqs.~(\ref{RiemannInv}),
the relationship between the variables $\rho$ and $u$.
Consequently, on the $(u,\rho)$-plane these simple wave solutions are
depicted as curves (see Fig.~\ref{Fig1})
\begin{equation}\label{}
    \begin{split}
    \rho=\frac{2+2\delta r_\pm + \delta u\pm2\sqrt{(1+2\delta r_\pm)(1 + \delta u)}}{\delta^2}.
    \end{split}
\end{equation}
To have the
intensity positive, it is necessary to fulfill the condition $-1/(2\delta)\leq r_-\leq r_+$.
Both curves touch the boundary line $u=-1/\delta$ of the instability region.
In Fig.~\ref{Fig1}, the modulationally unstable region (\ref{ModulUnstable}) is gray.
Along the line $\rho=(1+\delta u)/\delta^2$ both derivatives $\prt r_+/\prt u=0$, $\prt r_+/\prt \rho=0$ vanish.
We say that this line
separates two monotonicity regions in the half-plane $\rho\geq0$ (see Fig.~\ref{Fig1}).
The two intersection points of curves correspond to uniform flows with
constant parameters $\rho=\mathrm{const}$ and $u=\mathrm{const}$, that is to
the plateau solutions.
It is easy to express the physical variables $\rho$ and $u$ in terms of $r_-$ and $r_+$,
\begin{equation}\label{IandUbyR}
    \begin{split}
    \rho & = \frac{1}{2\delta^2} \left(1+\delta(r_++r_-)\pm\sqrt{(1+2\delta r_+)(1+2\delta r_-)}\right), \\
    u & = \frac{1}{2\delta}\left(\delta(r_++r_-)-1\mp\sqrt{(1+2\delta r_+)(1+2\delta r_-)}\right).
    \end{split}
\end{equation}

The initial profiles (\ref{init}), being infinitely sharp, do not
involve any characteristic length. Therefore
the large-scale features of the solution of this problem can
depend on the self-similar variable $\zeta=x/t$ only,
that is, $r_{\pm}=r_{\pm}(\zeta)$, and then the system (\ref{RiemannEquat}) reduces to
\begin{equation}\label{withselfsim}
    \begin{split}
    \left(v_--\zeta\right)\frac{dr_-}{d\zeta}=0, \quad \left(v_+-\zeta\right)\frac{dr_+}{d\zeta}=0.
    \end{split}
\end{equation}
We note again that these equations have a simple
solution $r_-=\mathrm{const}$,
$r_+=\mathrm{const}$ with constant $u$ and $\rho$, which corresponds
to the above-mentioned plateau region.

\begin{figure}[t] \centering
\includegraphics[width=0.45\textwidth]{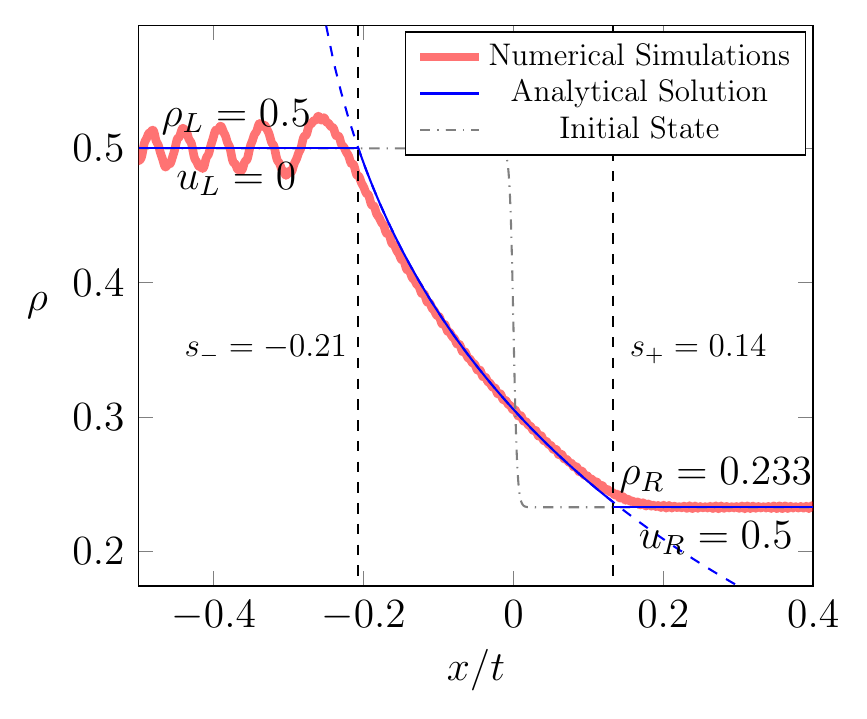}
\caption{Dependence of simple wave solution $\rho(\zeta)$ on $\zeta=x/t$ composed by a rarefaction wave connecting two uniform
flows. Numerical solution is shown in red (thick) and analytical solution is shown in blue (thin).
Vertical dashed lines indicate the edges of the rarefaction wave according to (\ref{VelRws}).
Gray dash-dotted line represents the initial state.
Here $\rho^L=0.5$, $u^L=5$, $\rho^R=0.233$, $u^R=0.5$, with $\delta=1$.}
\label{Fig2}
\end{figure}

Turning to self-similar simple wave solutions, let us consider for definiteness
the case when $r_{-}=\mathrm{const}$. Then $r_+$ changes in such a way
that the term between parentheses in the right equation (\ref{withselfsim}) is zero ($v_+=\zeta$),
so we have
\begin{equation}\label{I_tau}
    \begin{split}
    \rho(\zeta) & = \frac{1}{2\delta^2}\left\{ 1+\frac23\delta r_-+ \frac23\delta\zeta \right. \\
    & \left. \pm\sqrt{\left(1+2\delta r_-\right)\left(1-\frac23\delta r_-+\frac43\delta \zeta\right)} \right\},\\
    u(\zeta) & = \frac23(r_-+\zeta)-\delta \rho(\zeta).
    \end{split}
\end{equation}
We see that in the self-similar solutions the variable $\zeta$ must be above its minimal value
\begin{equation}\label{}
    \begin{split}
    \zeta\geq\frac{r_-}{2}-\frac{3}{4\delta}.
    \end{split}
\end{equation}
Similar formulas and plots can be obtained for the solution
$r_+=\mathrm{const}$, $v_-(r_-,r_+)=x/t \equiv \zeta$.
This wave configuration represents a rarefaction wave.
In the general case this type of wave can connect uniform
flows with equal values of the corresponding Riemann invariants
$r_-^L=r_-^R$ or $r_+^L=r_+^R$. An example of the corresponding
distribution is shown in Fig.~\ref{Fig2}. The analytical simple
wave approximation (thin blue) agrees with the numerical
solution of the generalized CLL equation (\ref{gCLL}) (thick red)
very well.

\begin{figure}[t] \centering
\includegraphics[width=0.5\textwidth]{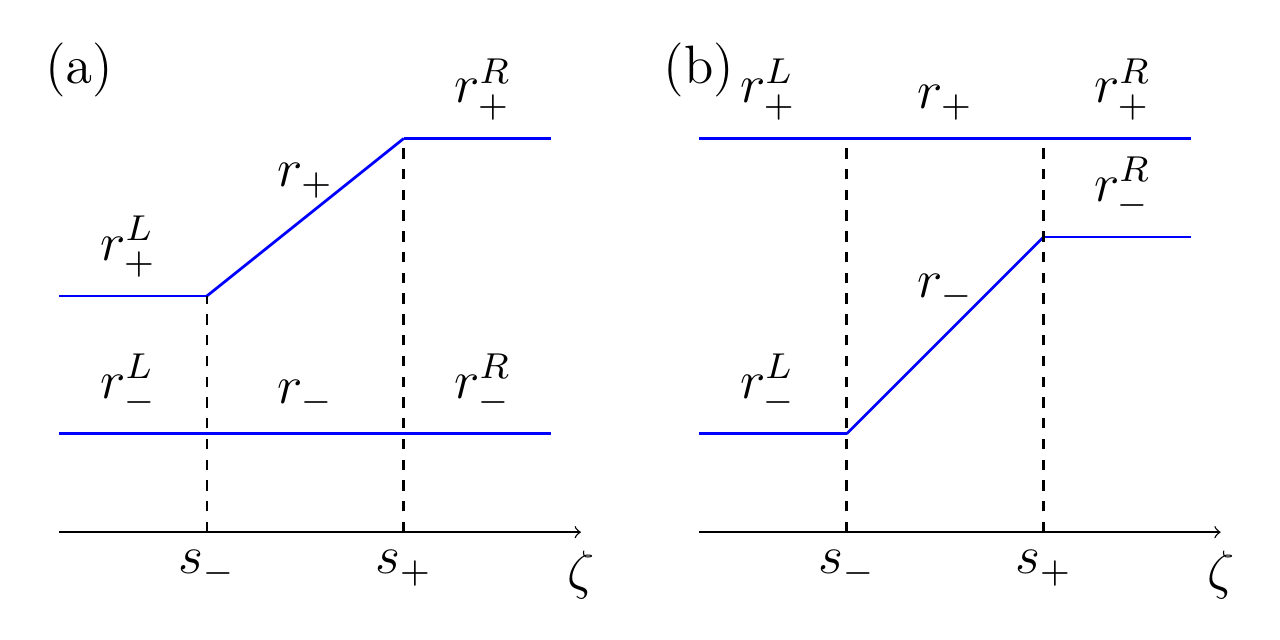}
\caption{Diagrams representing the evolution of the Riemann invariants as a function of $\zeta=x/t$ for two types of rarefaction waves.}
\label{Fig3}
\end{figure}

It is clear that in these self-similar solutions one of the Riemann invariants
must be constant and another one must increase with $\zeta$
according to
\begin{equation}\label{300.7}
    r_-=r_-^0=\mathrm{const},\qquad {v_+}=\frac32r_++\frac12r_-^0=\zeta
\end{equation}
or
\begin{equation}\label{300.71}
    r_+=r_+^0=\mathrm{const},\qquad {v_-}=\frac32r_-+\frac12r_+^0=\zeta.
\end{equation}
The dependence of the Riemann
invariants on the physical parameters must also be
monotonic in order to keep the solution single-valued.
The dependence of the Riemann invariants on $\zeta$ is sketched in Fig.~\ref{Fig3}
for two possible situations with $r_-$ or $r_+$ constant.
The edge velocities of these rarefaction waves are equal to
\begin{equation}\label{VelRws}
    \begin{split}
    & (a) \quad s_-=\frac{1}{2}r_-^L+\frac{3}{2}r_+^L,
        \quad s_+=\frac{1}{2}r_-^R+\frac{3}{2}r_+^R;\\
    & (b) \quad s_-=\frac{3}{2}r_-^L+\frac{1}{2}r_+^L,
        \quad s_+=\frac{3}{2}r_-^R+\frac{1}{2}r_+^R.
    \end{split}
\end{equation}
Obviously, the corresponding wave structures must satisfy the conditions (a) $r_+^L<r_+^R$,
$r_-^L=r_-^R$ or (b) $r_+^L = r_+^R$, $r_-^L<r_-^R$.
The other two situations with opposite inequalities
result in
multi-valued solutions and are therefore non-physical: the
dispersionless approximation is not applicable to these cases and we
have to turn to another type of key elements for describing such
structures.

\subsection{Cnoidal dispersive shock waves}

\begin{figure}[t] \centering
\includegraphics[width=0.5\textwidth]{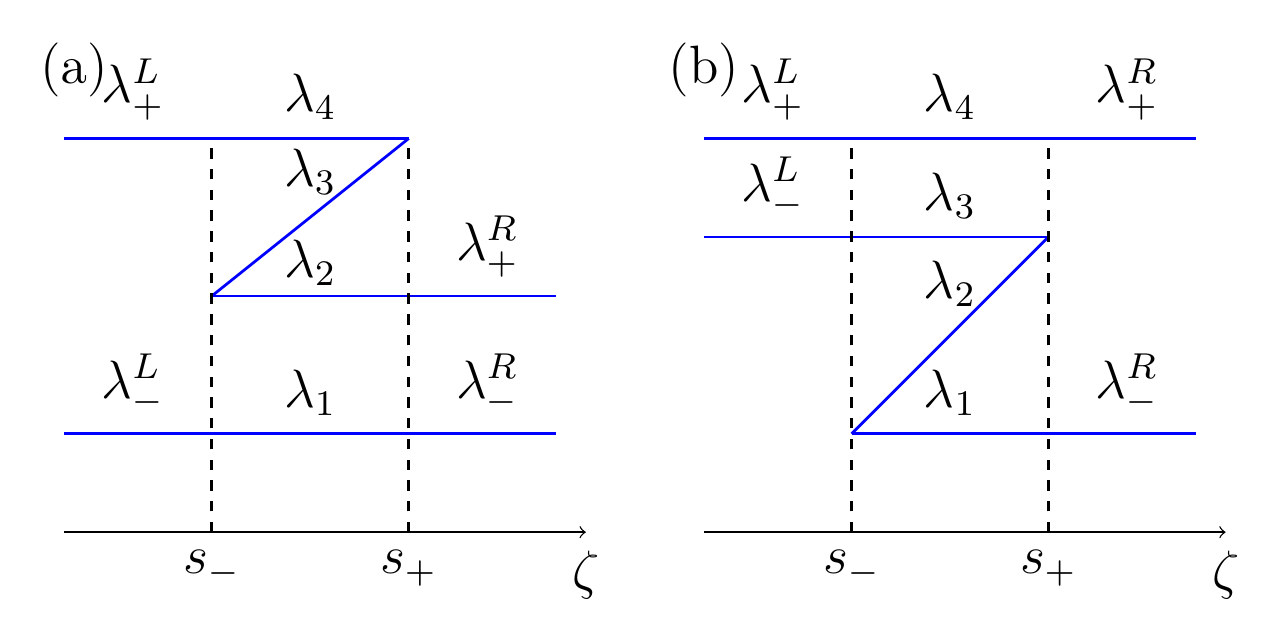}
\caption{Diagrams considered
within the dispersionless approximation correspond to a formal multi-valued solution. In this case, the dispersionless approximation breaks down and one observes a dispersive shock
wave, accurately described by four Riemann invariants within
the Whitham modulational approach.}
\label{Fig4}
\end{figure}

The other two possible solutions of Eqs.~(\ref{RiemannEquat}) are sketched
in Fig.~\ref{Fig4}, where for future convenience we have made
the change $r\mapsto\la$ ($r_\pm$ will be functions of $\la_\pm$ defined below),
and they satisfy the boundary conditions $(a)$ $\la_+^L = \la_+^R$,
$\la_-^L > \la_-^R$ or $(b)$ $\la_+^L > \la_+^R$, $\la_-^L = \la_-^R$.
We consider $\la_i$ as four
Riemann invariants of the Whitham system that describe evolution of a
modulated nonlinear periodic wave. We interpret this as a formation of the
cnoidal dispersive shock wave from the initial discontinuity with such a type
of the boundary conditions.

Since the pioneering work of Gurevich and Pitaevskii \cite{gp-1973}, it
has been known that wave breaking is
regularized by the replacement of the non-physical multi-valued
dispersionless solution by a dispersive shock wave. This wave pattern
can be represented approximately as a modulated nonlinear periodic
wave in which parameters $\la_i$ change slowly along the wave structure. In
this case, the two dispersionless Riemann invariants $\la_{\pm}$ (or $r_{\pm}$) are
replaced in the DSW region by four Riemann invariants $\la_i$.
In this region, the evolution of the DSW is
determined by the Whitham equations (\ref{WhithamEq}). If we
consider a self-similar solution, then all Riemann invariants depend only
on $\zeta=x/t$, and the Whitham equations reduce to
\begin{equation}\label{DispSelf-simEq}
    \left({v_i}-\zeta\right)\frac{d\la_i}{d\zeta}=0, \qquad i = 1,2,3,4.
\end{equation}
Hence we find again that only one Riemann invariant varies along the DSW, while
the other three are constant, that is the corresponding diagram reproduces the picture shown in
Fig.~\ref{Fig4}. The limiting expressions (\ref{sol-limit}) for
the Whitham velocities must coincide with expressions (\ref{riemann-vel}) for dispersionless
Riemann velocities and therefore we can relate the corresponding dispersionless and dispersive
Riemann invariants by the formulas
\begin{equation}\label{}
    \begin{split}
    (a) & \quad \lambda_-^L =\sqrt{1+2\delta r_-}, \quad
         \lambda_+^L=\sqrt{1+2\delta r_+}, \\
    (b) & \quad \lambda_-^R  =\sqrt{1+2\delta r_-}, \quad
         \lambda_+^R=\sqrt{1+2\delta r_+}
    \end{split}
\end{equation}
at the soliton edges of the DSW.
Here ${r}_{\pm}^{L,R}$ are the Riemann invariants of the dispersionless
theory that are defined by Eqs.~(\ref{RiemannInv}). They describe
the plateau solution at the soliton edge of the DSW. In a similar way, at the small-amplitude
edges we find similar relations
\begin{equation}\label{}
    \begin{split}
    (a) \quad \lambda_-^R & =\sqrt{1+2\delta r_-}, \quad
    \lambda_+^R=\sqrt{1+2\delta r_+},
    \end{split}
\end{equation}
and
\begin{equation}\label{}
   \begin{split}
    (b) \quad \lambda_-^L & =\sqrt{1+2\delta r_-}, \quad
    \lambda_+^L=\sqrt{1+2\delta r_+}.
    \end{split}
\end{equation}
Again the limiting expressions (\ref{small-limit}) and (\ref{small-limit2})
coincide with the dispersionless expressions (\ref{riemann-vel}).
Then the self-similar solutions of the Whitham equations (\ref{DispSelf-simEq})
are given by
\begin{equation}\label{RInv-sols}
    \begin{split}
    &(a) \quad v_3(\la_-^L,\la_+^R ,\la_3(\zeta),\la_+^L)=\zeta\; ;\\
    &\mbox{or}\\
    &(b) \quad v_2(\la_-^R,\la_2(\zeta),\la_-^L,\la_+^L)=\zeta\; ,
  \end{split}
\end{equation}
which define the dependence of the Riemann invariants (modulation parameters) $\la_3$ or $\la_2$
on $\zeta$ in implicit form.
The edges of the DSW propagate with velocities
\begin{equation}\label{VelDSWEdges}
    \begin{split}
    (a) & \quad s_-=-\frac{1}{\delta}+\frac{1}{4\delta}\left((\lambda_-^{L})^2+2(\lambda_+^{R})^2+(\lambda_+^{L})^2\right)\; , \\
        & \quad s_+=-\frac{1}{\delta}+\frac{1}{\delta}(\lambda_+^{L})^2+\frac{1}{4\delta}
            \frac{\left((\lambda_+^R)^2-(\lambda_-^R)^2\right)^2}{(\lambda_+^R)^2+(\lambda_-^R)^2-2(\lambda_+^L)^2}\; ;\\
    (b) & \quad s_-=-\frac{1}{\delta}+\frac{1}{\delta}(\lambda_+^R)^2+\frac{1}{4\delta}
        \frac{\left((\lambda_+^L)^2-(\lambda_-^L)^2\right)^2}{(\lambda_+^L)^2+(\lambda_-^L)^2-2(\lambda_+^R)^2}\; ,\\
        & \quad s_+=-\frac{1}{\delta}+\frac{1}{4\delta}\left((\lambda_-^R)^2+2(\lambda_+^L)^2+(\lambda_+^R)^2\right)\; .
    \end{split}
\end{equation}

\begin{figure}[t] \centering
\includegraphics[width=0.45\textwidth]{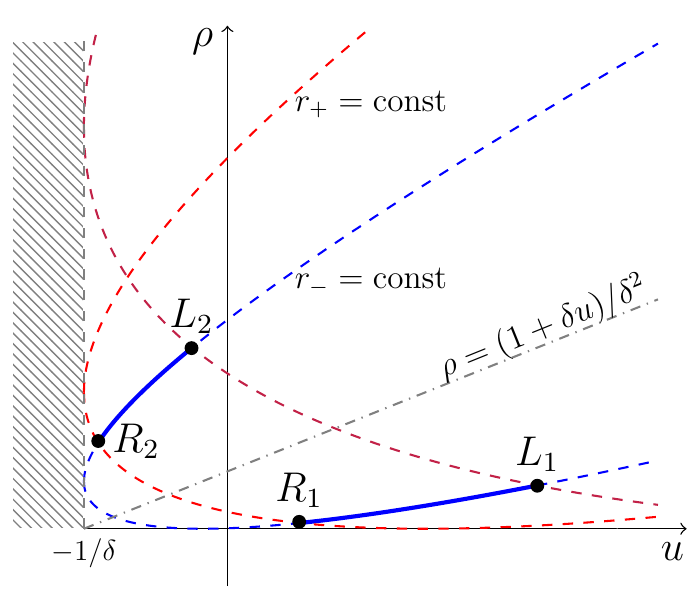}
\caption{Example of two possible paths in the ($u,\rho$)-plane between the left and right
boundary for the case of dispersive shock waves.
Corresponding wave structures are shown in Fig.~\ref{Fig6} and they satisfy the same solution
of the Whitham equations, but different boundary conditions in physical variables.}
\label{Fig5}
\end{figure}

\begin{figure}[t] \centering
    \includegraphics[width=0.45\textwidth]{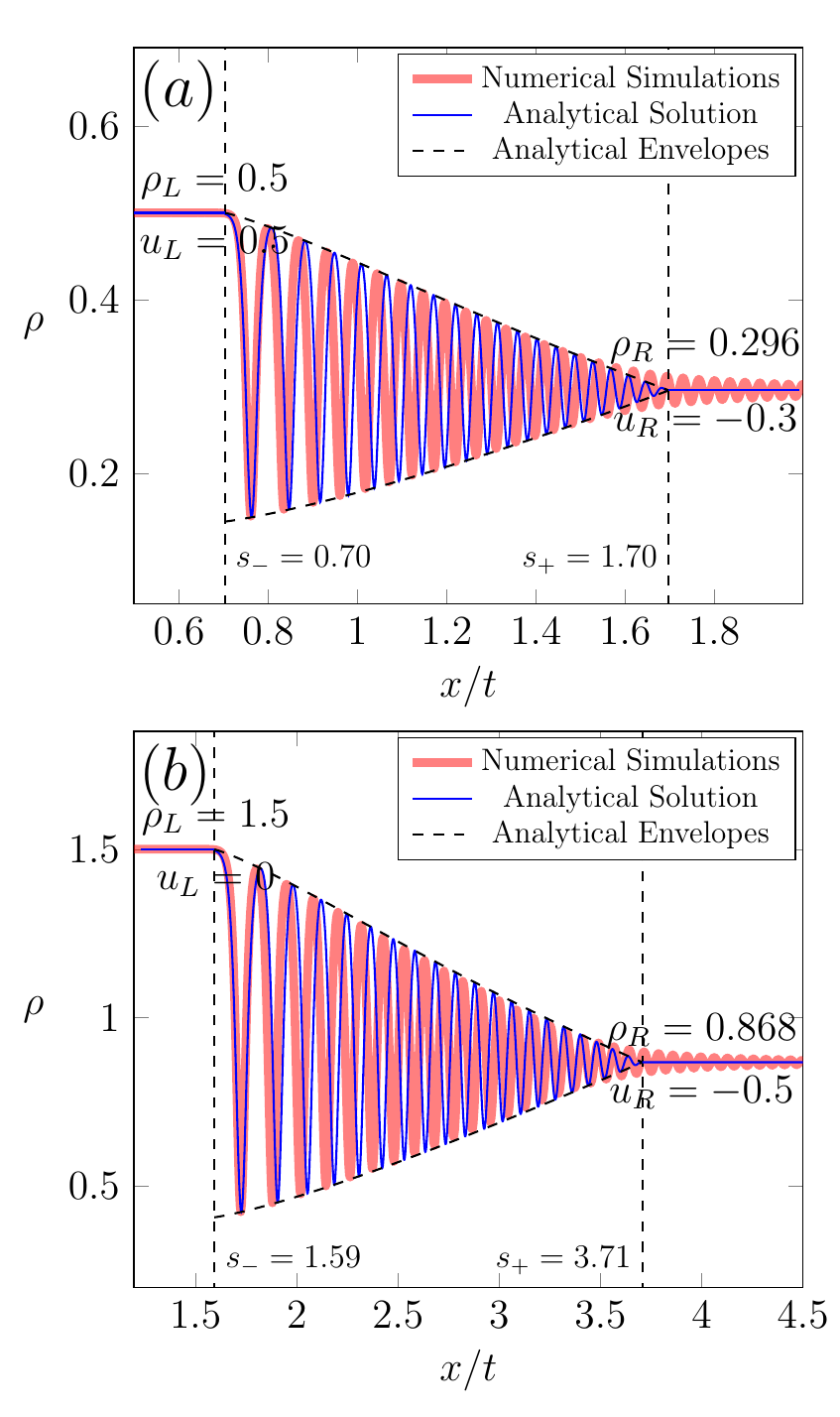}
    \caption{The comparison of analytical (thin blue) and numerical (thick red) solutions of the generalized CLL equation (\ref{gCLL})
    for two different boundary conditions and the same solution of the Whitham equations for
    the modulation parameters: (a) $\rho^L=0.5$, $u^L=0.5$, $\rho^R=0.296$, $u^R=-0.3$;
    (b) $\rho^L=1.5$, $u^L=0$, $\rho^R=0.868$, $u^R=-0.5$ with $\delta=1$.
    Dashed lines show analytical envelopes. Vertical dashed lines indicate the edges of the cnoidal DSW wave according to (\ref{VelDSWEdges}).}
    \label{Fig6}
\end{figure}

As shown in the diagrams of Fig.~\ref{Fig4}, three of the four Riemann invariants $\lambda_i$ in the DSW
are equal to the values of Riemann invariants on the boundaries. Moreover, the third invariant changes
according to (\ref{RInv-sols}). The substitution of $\lambda_i$
determines the dependence of $\nu_i$ on $\zeta$ for each of the cases in (\ref{zeros1}) and (\ref{zeros2}).
This means that there are two mappings from
Riemann invariants to the physical parameters. This point will be important in classification of
the wave structures evolved from the initial discontinuities.
For example, let us consider the case (a) ($\la_-^L=\la_-^R$, $\la_+^L>\la_+^R$).
We have two paths on the $(u,I)$-plane that satisfy this choice.
These two paths $L_1 \rightarrow R_1$ and $L_2 \rightarrow R_2$
are shown in Fig.~\ref{Fig5} and correspond to two mappings (\ref{zeros1}) and (\ref{zeros2}),
where points $L_1$ and $L_2$ correspond to the left boundary
condition with the Riemann invariants equal to $\la_-^L$ and $\la_+^L$,
and the points $R_1$ and $R_2$ correspond to the right
boundary condition with the Riemann invariants equal to $\la_-^R$ and $\la_+^R$.
In Fig.~\ref{Fig6} we compare
the analytic solution in the Whitham approximation with the exact
numerical solution of the generalized CLL equation. One can see that the envelope functions resulting from
the Whitham approach (dashed black lines) agree very well with the exact
numerical solution (thick red lines).

In a similar way, the diagram Fig.~\ref{Fig4}(b)
produces two other wave structures.

\subsection{Contact dispersive shock waves}

We now turn to the study of the situation where the
left and right boundary points belong to different monotonicity regions.
First, we consider the situation in which the Riemann
invariants have equal values at both edges of the shock, i.e., when
$r_-^L=r_-^R$, $r_+^L=r_+^R$ and, consequently, $\la_-^L=\la_-^R$, $\la_+^L=\la_+^R$.
This situation resembles the one of
the so called `contact discontinuities' which play an important role
in the theory of viscous shocks (see, e.g., Ref. \cite{LandauLifshitz-59}); therefore we
shall denote the wave structures arising in this case as {\it contact
dispersive shock waves} (to avoid any confusion, we should
mention that in the dynamics of immiscible condensates, interfaces
between two components may appear which play the same role as the one
played by contact discontinuities in the theory of viscous shocks;
see, e.g., \cite{IvanovKamchatnov-17}).
{{This type of DSW was first reported in \cite{Marchant-08} where
	the evolution of a step problem was studied for the focusing mKdV
	equation (see also a similar solution for the
	complex modified mKdV equation in \cite{KPT-08}).
	In \cite{EHS-17} these (trigonometric) DSWs were first attributed the name of contact DSWs.
	The contact DSWs of the generalized CLL equation are described
	by the modulated finite-amplitude nonlinear
	periodic solutions (\ref{eq26}) or (\ref{eq33}). At one of the edges
	of the trigonometric shock
	the amplitude vanishes and at the opposite edge
	it assumes some finite value. Generically, as will be explained
	later, contact DSWs are realized as parts of
	composite solutions (either a combination of cnoidal and
	trigonometric shocks or a combination of a trigonometric
	DSW and a rarefaction wave).}}
Such contact waves can arise only if the boundary points are located on the opposite
sides of the line $\rho=(1+\delta u)/\delta^2$, i.e., in different regions of monotonicity.
{{The diagram shown in Fig.~\ref{Fig7} corresponds to the path in Fig.~\ref{Fig8}.
	 In this case, the curve connecting the end points crosses the line $\rho=(1+\delta u)/\delta^2$
	of the hyperbolicity square along which $\lambda_-$ takes its minimal value: $\lambda_- = 0$.
	This means that in the formal dispersionless solution, the invariant $\lambda_-$
	would first decrease and reach its minimal value,
	then increase to the initial value along the same `path'.
We see in Fig.~\ref{Fig7} that two Riemann invariants $\lambda_3$ and $\lambda_4$
are constant within the shock region and they match
the boundary condition $\lambda_3=\lambda_-^L=\lambda_-^R$, $\lambda_4=\lambda_+^L=\lambda_+^R$,
whereas the two other
Riemann invariants are equal to each other ($\la_1=\la_2$) and satisfy the same
Whitham equation with $v_1(\la_1,\la_1,\lambda_-^L,\lambda_+^L) = v_2(\la_1,\la_1,\lambda_-^L,\lambda_+^L)=\zeta$.
Here $\zeta$ varies within the interval $s_-\leq \zeta\leq s_+$ with
\begin{equation}\label{VelTDSWEdges}
\begin{split}
    s_-&=-\frac{1}{\delta}-\frac{1}{4\delta}\frac{\left((\lambda^L_+)^2-(\lambda^L_-)^2\right)^2}{(\lambda^L_-)^2+(\lambda^L_+)^2}, \\
    s_+&=-\frac{1}{\delta}+\frac{1}{4\delta}\left(3(\lambda^R_-)^2+(\lambda^R_+)^2\right).
    \end{split}
\end{equation}

\begin{figure}[t] \centering
\includegraphics[width=4.5cm]{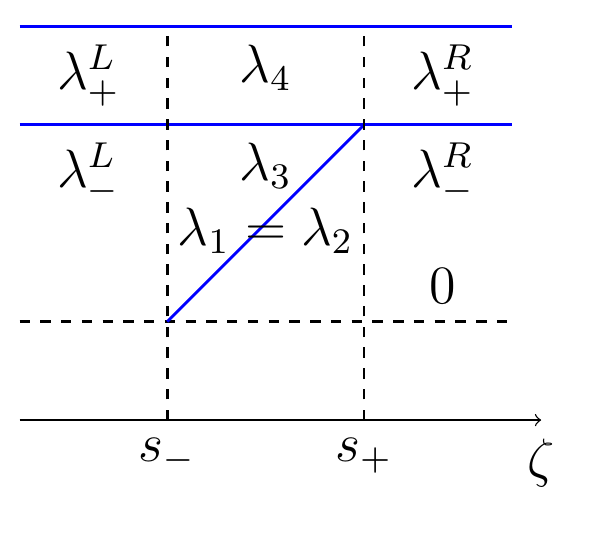}
\caption{Diagram represents evolution of the Riemann invariants as functions of $\zeta=x/t$
in the contact DSW solution of the Whitham equations: $r_-^L=r_-^R$ ($\lambda_-^L=\lambda_-^R$),
$r_+^L=r_+^R$ ($\lambda_+^L=\lambda_+^R$).}
\label{Fig7}
\end{figure}

\begin{figure}[t] \centering
\includegraphics[width=0.45\textwidth]{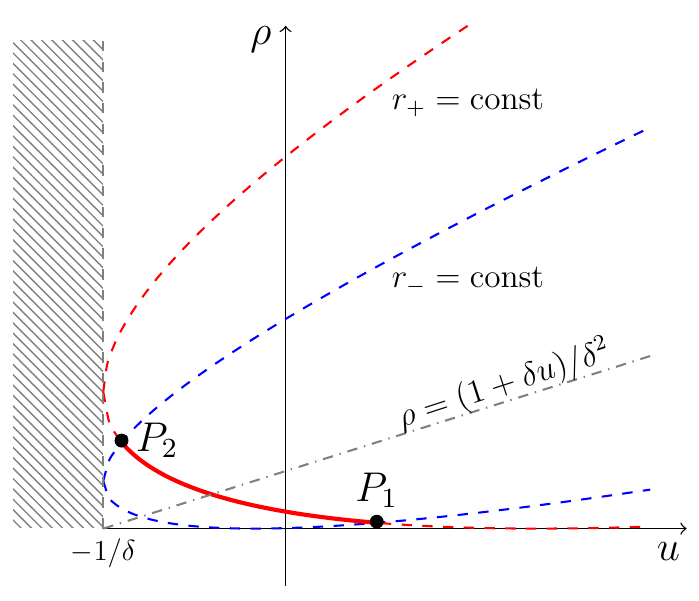}
\caption{Example of path for a contact DSW ($r_+^L=r_+^R$ and $r_-^L=r_-^R$) with crossing of $\rho=(1+\delta u)/\delta^2$ line separating the regions of monotonicity.
Two directions $P_1 \leftrightarrow P_2$ are described by the mappings (\ref{zeros1}) and
(\ref{zeros2}).}
\label{Fig8}
\end{figure}

As in the case of cnoidal DSWs, due to different mappings
(\ref{zeros1}) or (\ref{zeros2}) the single contact diagram
corresponds to two DSW structures.
An example of two paths $P_1 \rightarrow P_2$ and the opposite $P_2 \leftarrow P_1$
is shown in Fig.~\ref{Fig8}. The corresponding wave structures are shown in Fig.~\ref{Fig9}.
For the contact DSW the wave amplitude varies in a quadratic manner through the shock. This is particularly noticeable near the small-amplitude edge of the DSW. This is in contrast to the cnoidal DSW for which the wave amplitude varies linearly.

\begin{figure}[t] \centering
\includegraphics[width=0.45\textwidth]{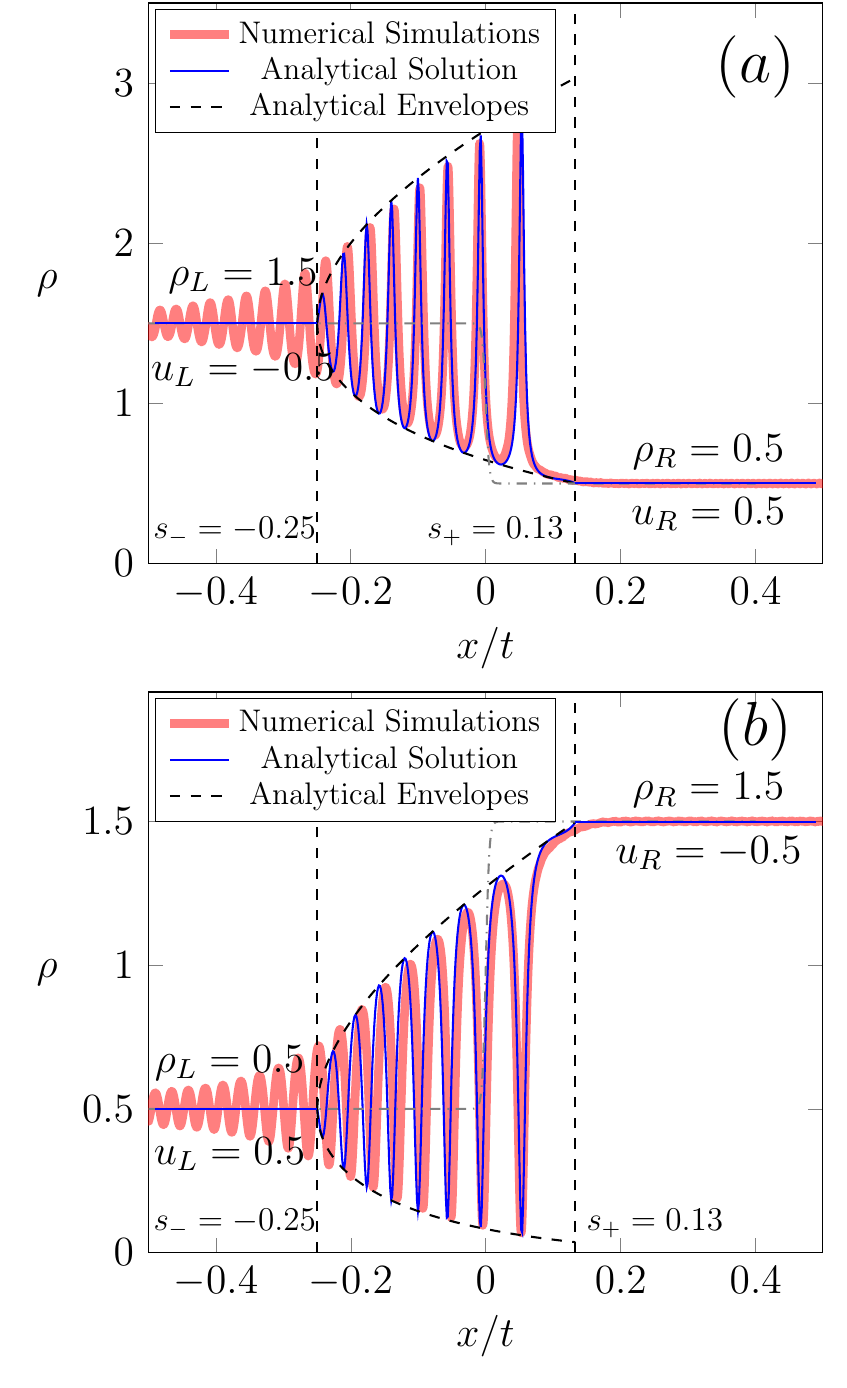}
\caption{The comparison of analytical (thin blue) and numerical (thick red) solutions of the generalized CLL equation (\ref{gCLL})
for contact DSW for two possible choices of directions $P_1\longleftrightarrow P_2$ and
corresponding mappings (\ref{zeros1}) and (\ref{zeros2}).
Here (a) $\rho^L=1.5$, $u^L=-0.5$, $\rho^R=0.5$, $u^R=0.5$;
(b) $\rho^L=0.5$, $u^L=0.5$, $\rho^R=1.5$, $u^R=-0.5$ and $\delta=1$.
Dashed black lines show analytical envelopes; gray dash-dotted line represents the initial state.
Vertical dashed lines indicate the edges of the contact DSW wave according to (\ref{VelTDSWEdges}).}
\label{Fig9}
\end{figure}

\subsection{Combined shocks}

It is natural to ask what happens if one of the Riemann invariants still remains constant
($r_+^L=r_+^R$ or $\la_+^L=\la_+^R$); however the boundary values of the other Riemann invariant are different:
$r_-^L<r_-^R$ ($\la_-^L>\la_-^R$) or $r_-^L>r_-^R$ ($\la_-^L<\la_-^R$)).
To be definite, we shall consider two generalizations of the situation.
The transition of the type $L_1\to R_1$ or $L_2\to R_2$ of
Fig.~\ref{Fig8} can be generalized in two ways represented in
Fig.~\ref{Fig10}, where the points $L_i$ and $R_i$ symbolize plateaus at
the left and right boundaries, respectively.
In this case the boundary points are also
located in different monotonicity regions.
Diagrams for Riemann invariants for both of these cases are
shown in Fig.~\ref{Fig11}. 

\begin{figure}[t] \centering
\includegraphics[width=0.4\textwidth]{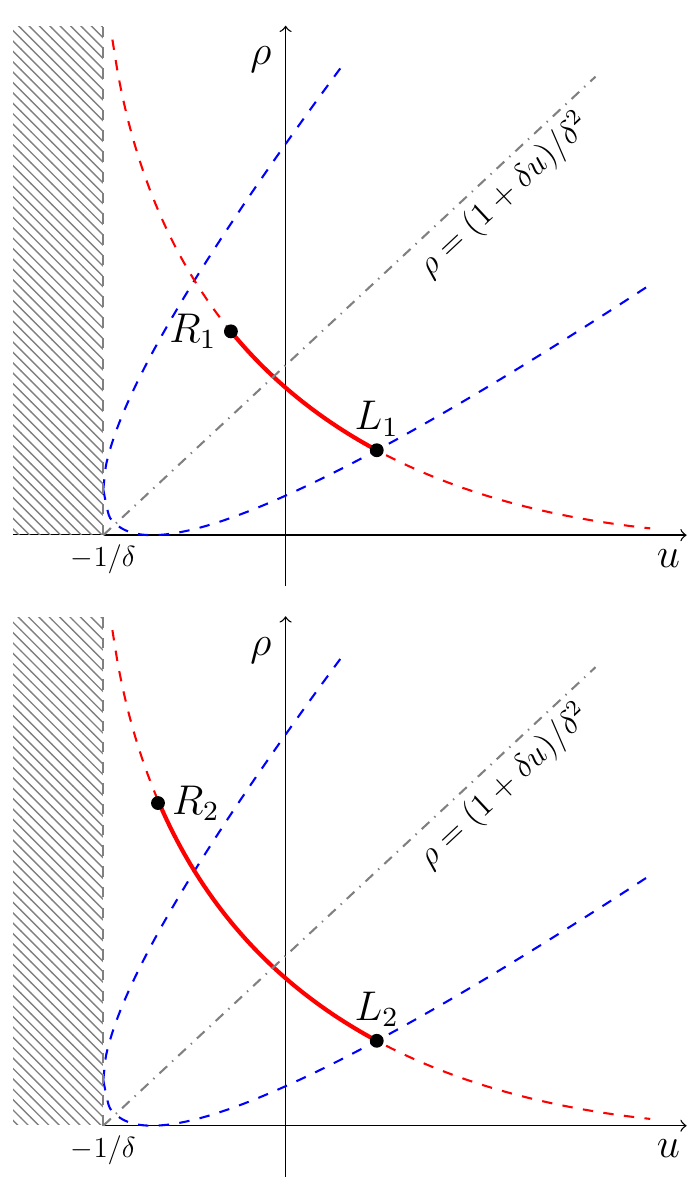}
\caption{Paths in the $(u,I)$-plane associated with two types of combined shocks.
The left and right boundary conditions correspond to points $L$ and $R$, respectively;
they lie on the curves along which the dispersionless Riemann invariant $r_+=r_+^L=r_+^R$ ($\la_+=\la_+^L=\la_+^R$) is constant.
One has $r_-^L<r_-^R$ ($\la_-^L>\la_-^R$) in case (a) and $r_-^L>r_-^R$ ($\la_-^L<\la_-^R$) in case (b).}
\label{Fig10}
\end{figure}

\begin{figure}[t] \centering
	\includegraphics[width=0.5\textwidth]{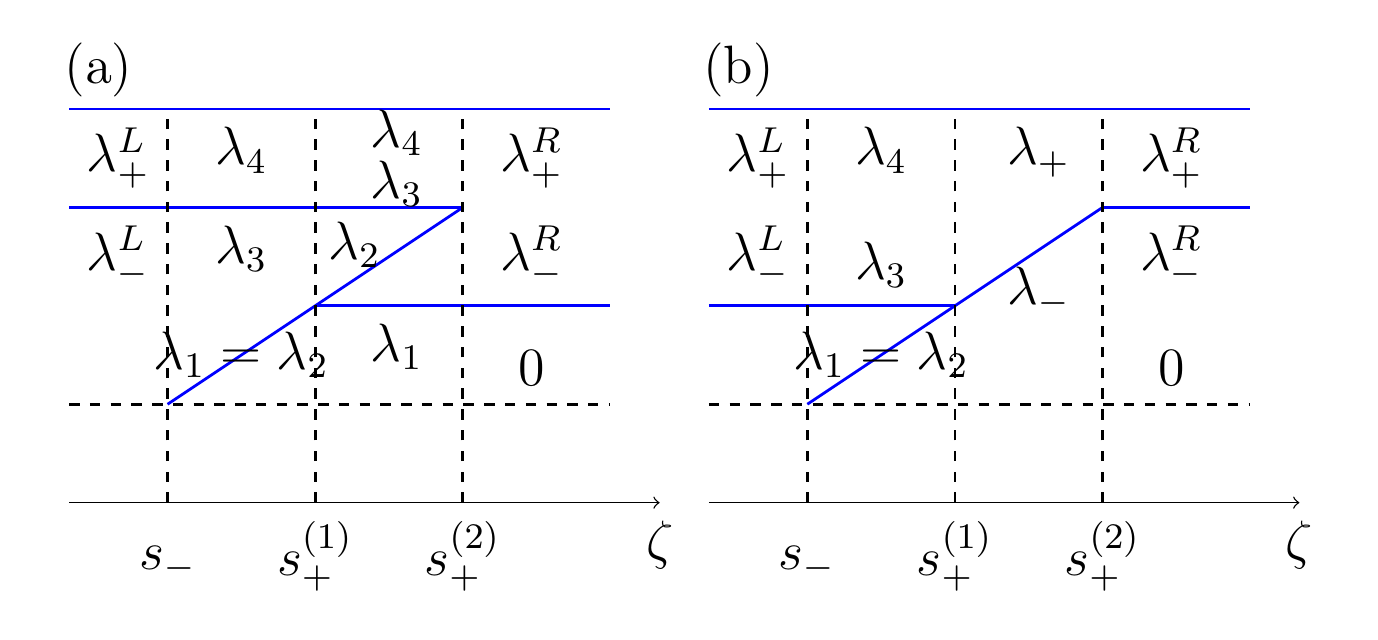}
	\caption{Diagrams representing the evolution of the Riemann invariants as functions
		of $\zeta=x/t$ for combined shocks corresponding to the paths in the $(u,\rho)$-plane shown in Fig.~\ref{Fig10}.}
	\label{Fig11}
\end{figure}

In the case corresponding to Fig.~\ref{Fig10}(a) the contact trigonometric
DSW is attached at its right edge to the cnoidal DSW.
At the right soliton edge the cnoidal wave matches the right
boundary plateau. The velocities of the characteristic points
identified in Fig.~\ref{Fig11}(a) are given by
\begin{equation}\label{CombDSWEdges1}
    \begin{split}
    & s_{-}=-\frac1{\delta} - \frac{1}{4\delta}\frac{\left((\lambda_+^L)^2-(\lambda_-^L)^2\right)^2}{(\lambda_-^L)^2+(\lambda_+^L)^2}, \\
    & s_{+}^{(1)}=-\frac1{\delta} + \frac{1}{4\delta}\left(3(\lambda_+^L)^2+(\lambda_-^L)^2\right),\\
    & s_+^{(2)}=-\frac1{\delta} + \frac{1}{4\delta}\left((\lambda_-^R)^2+2(\lambda_-^L)^2+(\lambda_+^R)^2\right).
    \end{split}
\end{equation}
The resulting composite wave structure is shown in Fig.~\ref{Fig12}(a)
(thin blue line) where it is compared with the numerical solution of the
generalized CLL equation (\ref{gCLL}) (thick red line).
	{{As an example, we take the boundary parameters $\rho^L=1$, $u^L=1$, $\rho^R=1.739$, $u^R=0.2$. For this parameter choice the  solution consists of a combination of the
	cnoidal DSW (\ref{eq30}) and part of the trigonometric DSW (\ref{eq33}). The leading portion of the DSW consists of a modulated cnoidal wave. At the leading (soliton) edge, at $s_+^{(2)}=0.55$, the modulus squared is $m=1$, so solitons occur there. At the trailing edge of the cnoidal DSW, at $s_+^{(1)}=0.40$,
	the modulus squared is $m=0$, and trigonometric waves occur.
	The rear portion of the numerically realized shock consists of part of the contact DSW. It
	extends from its small-amplitude edge, at $s_-=0.33$, where linear waves occur, to $s_+^{(1)}=0.40$, where it matches the trailing edge of the cnoidal DSW.
	There is a discontinuity in the slope of the theoretical envelope at the junction of these two different DSW types for combined solutions.
	The vertical dashed lines in the figure separate the parts of the composite wave.}}

\begin{figure}[t] \centering
\includegraphics[width=0.45\textwidth]{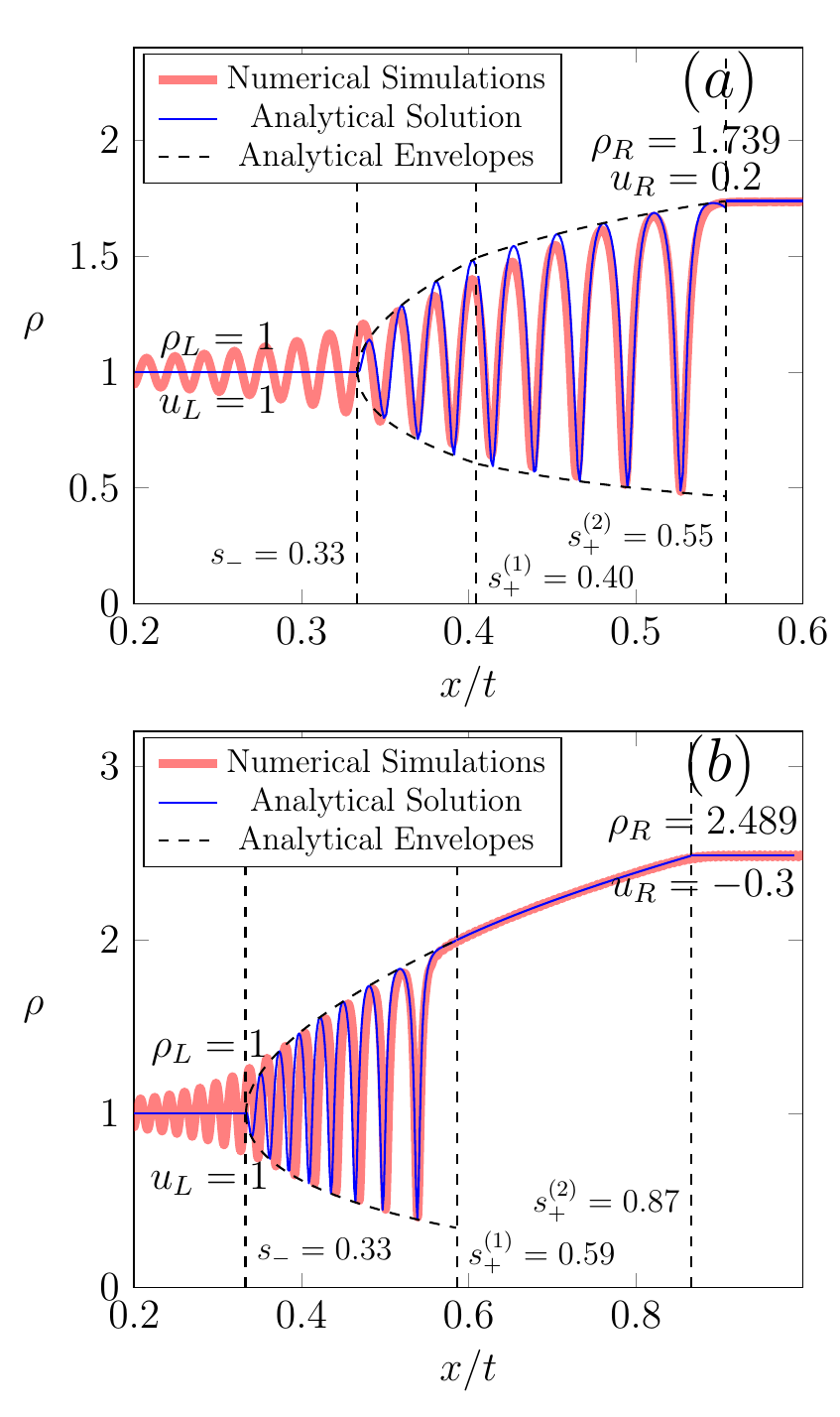}
\caption{The comparison of analytical (thin blue) and numerical (thick red) solutions of the generalized CLL equation (\ref{gCLL})
for combined shocks corresponding to the paths in the $(u,\rho)$-plane (Fig.~\ref{Fig10})
and to the diagrams of Riemann invariants (Fig.~\ref{Fig11}).
Here $\delta=1$ and (a) $\rho^L=1$, $u^L=1$, $\rho^R=1.739$, $u^R=0.2$; (b) $\rho^L=1$, $u^L=1$, $\rho^R=2.489$, $u^R=-0.3$.
Dashed black lines show analytical envelopes.
Vertical dashed lines indicate the edges of the combined DSW wave according to (\ref{CombDSWEdges1}) and (\ref{CombDSWEdges2}).}
\label{Fig12}
\end{figure}

In the case corresponding to Fig.~\ref{Fig10}(b) the contact DSW is attached
at its soliton edge to the rarefaction wave which matches at its right
edge the right boundary plateau. The velocities of the
characteristic points identified in Fig.~\ref{Fig11}(b)
are expressed in terms of the boundary
Riemann invariants by the formulas
\begin{equation}\label{CombDSWEdges2}
    \begin{split}
    & s_{-}=-\frac1{\delta} - \frac{1}{4\delta}\frac{\left((\lambda_+^L)^2-(\lambda_-^L)^2\right)^2}{(\lambda_-^L)^2+(\lambda_+^L)^2}, \\
    & s_{+}^{(1)}=-\frac1{\delta}+\frac{1}{4\delta}\left(3(\lambda_-^R)^2+(\lambda_+^L)^2\right), \\
    & s_+^{(2)}=-\frac1{\delta}+\frac{1}{4\delta}\left(3(\lambda_-^R)^2+(\lambda_+^R)^2\right).
    \end{split}
\end{equation}
The resulting combined wave structures are shown in Fig.~\ref{Fig12}(b)
(thin blue lines) where they are compared with the numerical solution of the
generalized CLL equation (thick red line).
	{{Here boundary parameters are $\rho^L=1$, $u^L=1$, $\rho^R=2.489$, $u^R=-0.3$.
	For this parameter choice the approximate solution consists of the trigonometric DSW (\ref{eq33})
	and the rarefaction wave of (\ref{I_tau}) type. The leading edge of the DSW, at $s_+^{(1)}=0.59$, consists of an algebraic soliton (\ref{eq34}). At the left edge of the DSW, at $s_-=0.33$, small-amplitude sinusoidal waves occur. The extent of rarefaction wave is $s_+^{(1)} < x/t < s_+^{(2)}$, where $s_+^{(1)}=0.59$ and $s_+^{(2)}=0.87$.
	One can see that agreement between numerical simulations and analytical results is very good.
	}}

This completes the characterization of all the key elements which may
appear in a complex wave structure evolving from an arbitrary initial
discontinuity of type (\ref{init}). We can now proceed to the
classification of all the possible composite structures.

\section{Classification of wave patterns}\label{sec5}

Now we turn to the Riemann problem. This problem arose long ago and it remains
an active area of research nowadays (see, e.g. \cite{Biondini-18,ENS-18,KWWDWX-19}). As was noted above, one of
the simplest cases is the KdV equation, where there are only two possible ways of evolution of initial
discontinuity: it can evolve into either a rarefaction wave or cnoidal DSW.
It was shown that the NLS equation evolution of any initial discontinuity
leads to a wave pattern consisting of a sequence of building blocks two of which are
represented by either the rarefaction wave or the DSW, and they are separated by
a plateau, or a vacuum, or a two-phase self-similar solution close to an unmodulated nonlinear
periodic wave.
In total, there are six different possible wave patterns that can evolve from a given
initial discontinuity. A similar classification of wave patterns was also established for the
dispersive shallow water Kaup-Boussinesq equation \cite{EGP-01,CIKP-17}.
For classification of wave patterns arising in solutions of the Riemann problem
of the KdV or NLS type, it is important that the corresponding dispersionless limits
are represented by the genuinely nonlinear
hyperbolic equations. If this is not the case, then the classification of the KdV-NLS
type becomes insufficient and it was found that it should include new elements ---
kinks, trigonometric dispersive shocks, or combined shocks. An example of such equations can be
the modified KdV \cite{Marchant-08}, Gardner \cite{kamch-2012}, or
Miyata-Camassa-Choi \cite{EslerPearce-08} equations.
These new elements can be labeled by two parameters only and
therefore these possibilities can be charted on a two-dimensional diagram.
In our present case the initial discontinuity (\ref{init}) is parametrized by
four parameters $u^L$, $\rho^L$, $u^R$, $\rho^R$; hence the number of possible wave patterns considerably
increases and it is impossible to present them in a two-dimensional chart.
Therefore it seems more effective to formulate the principles according to which
one can predict the wave pattern evolved from a discontinuity with given parameters.
Similar method was used \cite{ik-2017,ikcp-2017} in the classification of wave patterns
evolving from initial discontinuities according to the generalized NLS equation and
the Landau-Lifshitz equation.

\begin{figure}[t] \centering
\includegraphics[width=0.45\textwidth]{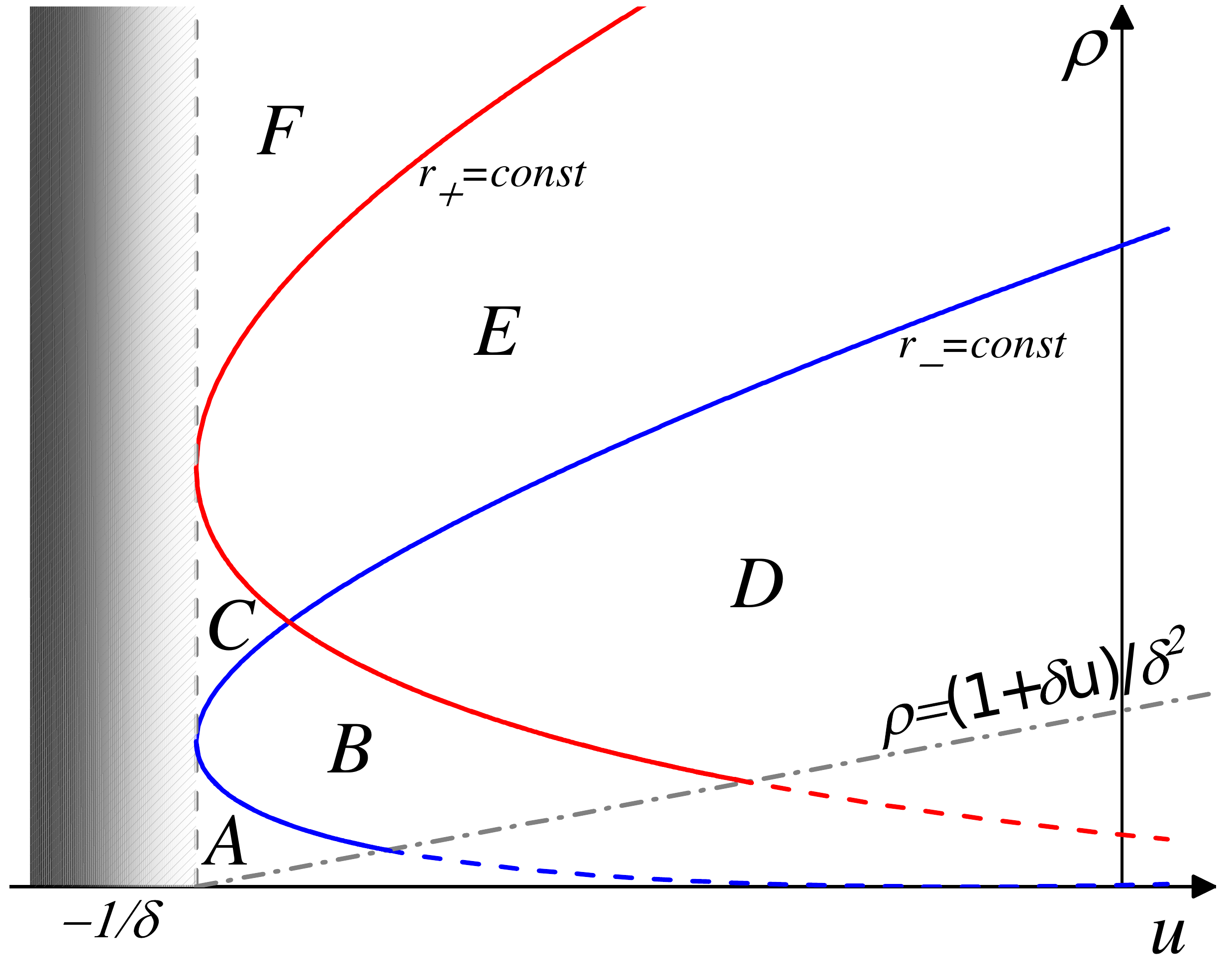}
\caption{Domains in the upper left monotonicity region of the ($u,\rho$)-plane
corresponding to different wave structures shown in Fig.~\ref{Fig14}.}
\label{Fig13}
\end{figure}

\begin{figure*}[ht] \centering
\includegraphics[width=1\textwidth]{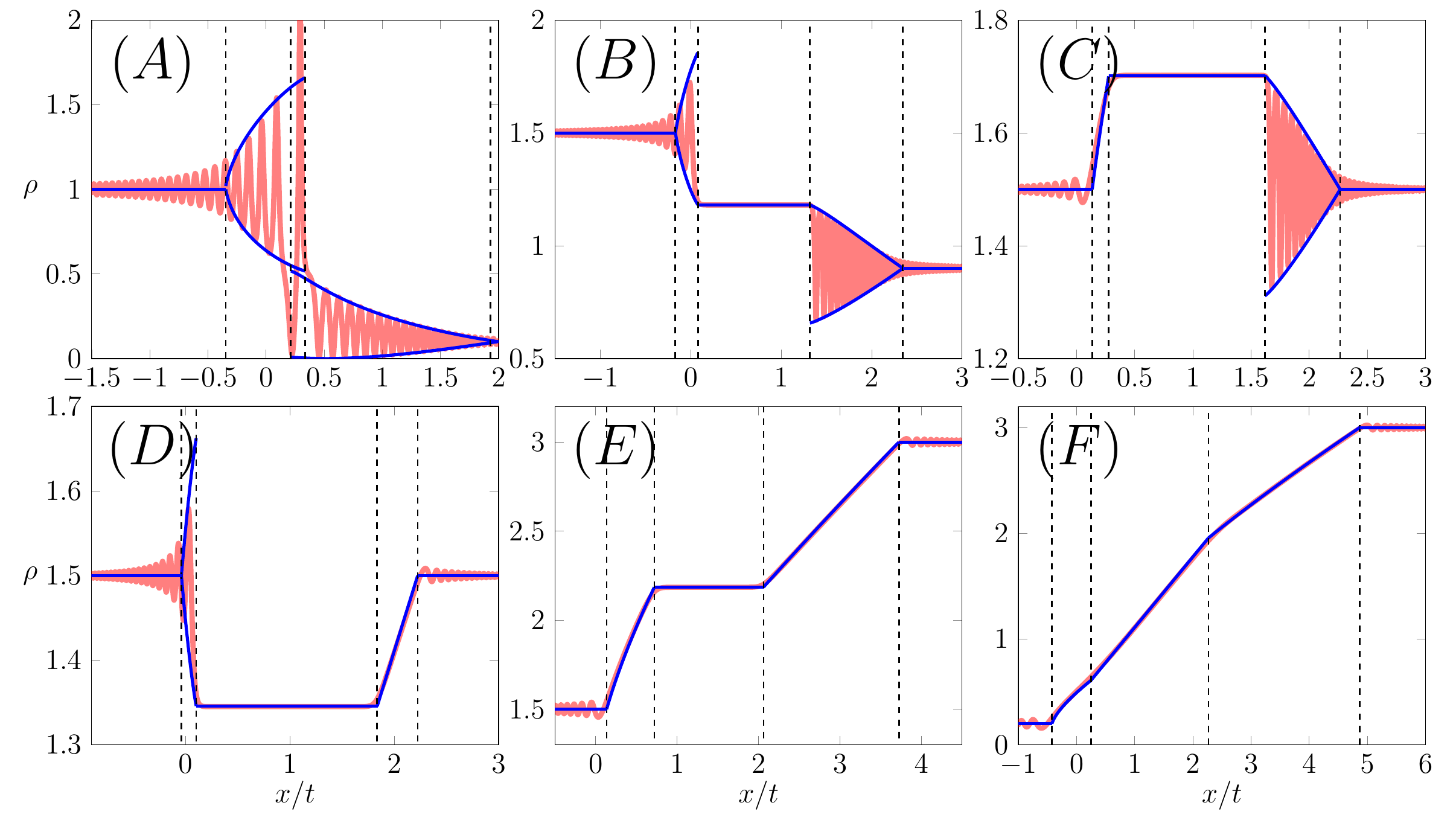}
\caption{Examples of wave structures corresponding to the location of
the point $R$ referring to the right boundary in one of the six
domains shown in Fig.~\ref{Fig13}.
In this case, the points corresponding to the boundaries of the initial state belong to one monotonicity region; therefore, the structures consist of a sequence of building blocks two of which are represented by either the rarefaction wave or the cnoidal DSW. Red (thick) curves show the numerical solution of the generalized CLL equation (\ref{gCLL}); blue (thin) curves illustrate the analytical solution. Vertical dashed lines reflect the edges of different waves.}
\label{Fig14}
\end{figure*}

As is clear from the previous section, it is convenient to distinguish
the situations where both points representing the left and right
boundary conditions belong to the same region of monotonicity from
those where they belong to different such regions (see Fig.~\ref{Fig13}).
It is convenient to begin with the consideration of the classification problem from the case
when both boundary points lie on one side of the line $\rho=(1+\delta u)/\delta^2$ separating two
monotonicity regions in the $(u,\rho)$-plane.
For definiteness, we denote the point of coordinates
$(u_L,\rho_L)$ referring to the left boundary by $L$ and plot the two
curves of constant Riemann invariants $\lambda_{+}^L$ and $\lambda_{-}^L$.
These divide the region into six sub-domains. It is easy to see
that, when the point $R$ referring to the right boundary is located in
one of these domains (labeled by the symbols A, B, ..., F), one of
the following inequalities is fulfilled:
\begin{equation}\label{RiemannInequalities}
    \begin{split}
    & \mbox{A}: \: \la_-^R < \la_+^R < \la_-^L < \la_+^L, \quad
       \mbox{B}: \: \la_-^R < \la_-^L < \la_+^R < \la_+^L, \\
    & \mbox{C}: \: \la_-^L < \la_-^R < \la_+^R < \la_+^L, \quad
        \mbox{D}: \: \la_-^R < \la_-^L < \la_+^L < \la_+^R, \\
    & \mbox{E}: \: \la_-^L < \la_-^R < \la_+^L < \la_+^R, \quad
        \mbox{F}: \: \la_-^L < \la_+^L < \la_-^R < \la_+^R.
  \end{split}
\end{equation}
The corresponding sketches of wave
structures are shown in Fig.~\ref{Fig14}.
In cases (B)--(E) two elementary wave structures presented in the previous section
are connected by a plateau whose
parameters are determined by the dispersionless Riemann invariants
$\lambda_{\pm}^P$ equal to $\lambda_-^P=\lambda_-^R$ and $\lambda_+^P=\lambda_+^L$.
In case (F) two rarefaction waves are separated by a region with intensity
and chirp which are expressed by formulas
\begin{equation} \label{InstPlateau}
    \begin{split}
    \rho = \frac{1+\delta\zeta}{\delta^2}, \qquad u = -\frac{1}{\delta}.
    \end{split}
\end{equation}
The last property is a feature of the generalized CLL equation (\ref{gCLL}),
since similar behavior for intensity (density for Bose-Einstein
condensates or depth for water waves) was not previously observed.

Let us look at each case separately:

$\bullet$ In case (A) two DSWs are produced with a nonlinear wave which
can be presented as a non-modulated cnoidal wave between them.
The volution of the wave structure
is shown in Fig.~\ref{Fig14}(A).

$\bullet$ In case (B) two DSWs are produced with a plateau between them.
Here we have a collision of two light fluids (see Fig.~\ref{Fig14}(B)).

$\bullet$ In case (C) we obtain a DSW on the right,
a rarefaction wave on the left, and a plateau in
between is produced (see Fig.~\ref{Fig14}(C)).

$\bullet$ In case (D) we get the same situation as in case (C),
but now the DSW and rarefaction wave exchange their places
(see Fig.~\ref{Fig14}(D)).

$\bullet$ In case (E) two rarefaction waves are connected by a plateau.
Here rarefaction waves are able now to provide enough flux of the light fluid to create
a plateau in the region between them (see Fig.~\ref{Fig14}(E)).

$\bullet$ In case (F) two rarefaction waves are combined into a single wave structure where
they are separated by a region with parameters varying according to (\ref{InstPlateau}).
This means that two light fluids flow in opposite directions with velocities so large that the
rarefaction waves are not able to form a plateau between them.
The sketch of the wave structure
is shown in Fig.~\ref{Fig14}(F).

Now we turn to consideration of the classification problem for the case
when both boundary points lie below and to the right of the line $\rho=(1+\delta u)/\delta^2$.
This situation is shown in Fig.~\ref{Fig15}.
We see that the curves divide again this right monotonicity region into
six domains. For this case the Riemann invariants can have
the same orderings (\ref{RiemannInequalities}) as in the previous case.
Depending on the location of the right boundary point in a certain domain, the corresponding
wave structure will be formed. For all cases
these structures coincide with those for the previous case.
The only difference from the previous situation is structure (F), where in this case
two rarefaction waves are combined into a single wave structure where
they are separated by an empty region.
This means that two light fluids flow in opposite directions with velocities so large that the
rarefaction waves are not able to fill in an empty region between them.

\begin{figure}[t] \centering
\includegraphics[width=0.45\textwidth]{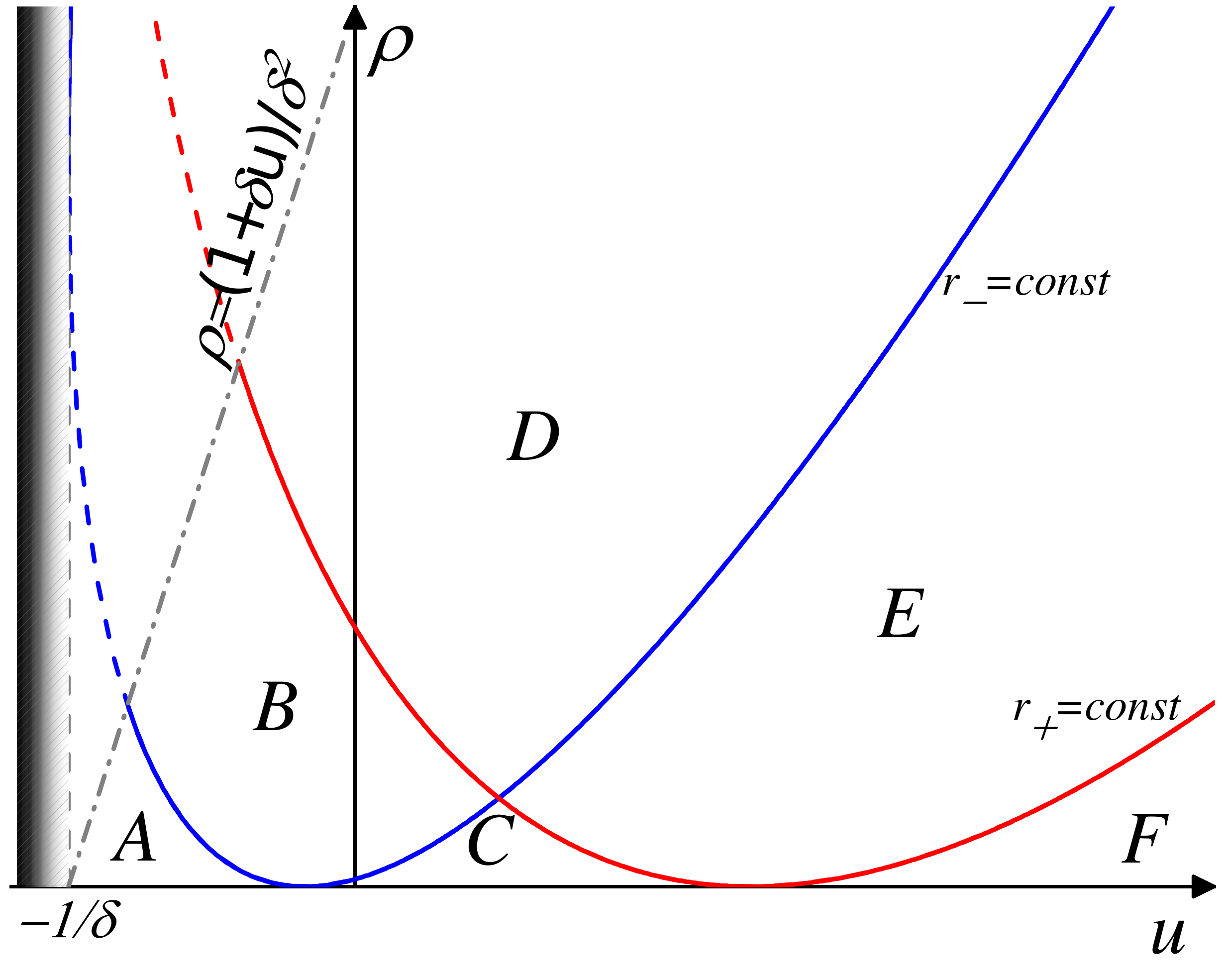}
\caption{Domains in the ($u,\rho$)-plane on the lower right side of the line $\rho=(1+\delta u)/\delta^2$ corresponding to different
structures.}
\label{Fig15}
\end{figure}

At last, we have to investigate the situation when the boundary points are located on different
sides of the line $\rho=(1+\delta u)/\delta^2$, that is, in different monotonicity regions.
As we have seen in the previous section, in this case new complex structures consisting
of contact DSWs or combined shocks appear.
Since the total number of possible wave patterns
is very large, we shall not list all of them here but
rather illustrate the general principles of their classification.

For given boundary parameters, we can construct the curves corresponding to
constant Riemann invariants $r_{\pm}^{L,R}$: each left or right pair of these
parabolas crosses at the point $L$ or $R$ representing the left or right
boundary state's plateau. Our task is to construct the path joining these two
points, then this path will represent the arising wave structure. We already know the answer
for the case when the left and right points lie on the same $r$-curve.
If this is not the case and the right point $R$ lies, say,
below the curve $r_-^L=\mathrm{const}$ then we can
reach $R$ by means of a more complicated path consisting of two curves
joined at the point $P$. Evidently, this point $P$ represents the plateau between two waves
represented by the curves. At the same time, each curve corresponds to a wave structure
discussed in the preceding section. In fact, there are
two paths with a single intersection point that join the left and right boundary points. We choose the
physically relevant path by imposing the condition that velocities of edges of all
regions must increase from left to right.
Having constructed a path from the left boundary point to the right one,
it is easy to draw the corresponding $\la$-diagram.
To construct the wave structure, we use
the formulas connecting the zeros $\nu_i$ of the resolvent with the Riemann
invariants $\lambda_i$ and expressions for the solutions parametrized by $\nu_i$.
This solves the problem of construction of
the wave structure evolving from the initial discontinuity with given boundary conditions.

\begin{figure}[t] \centering
\includegraphics[width=0.4\textwidth]{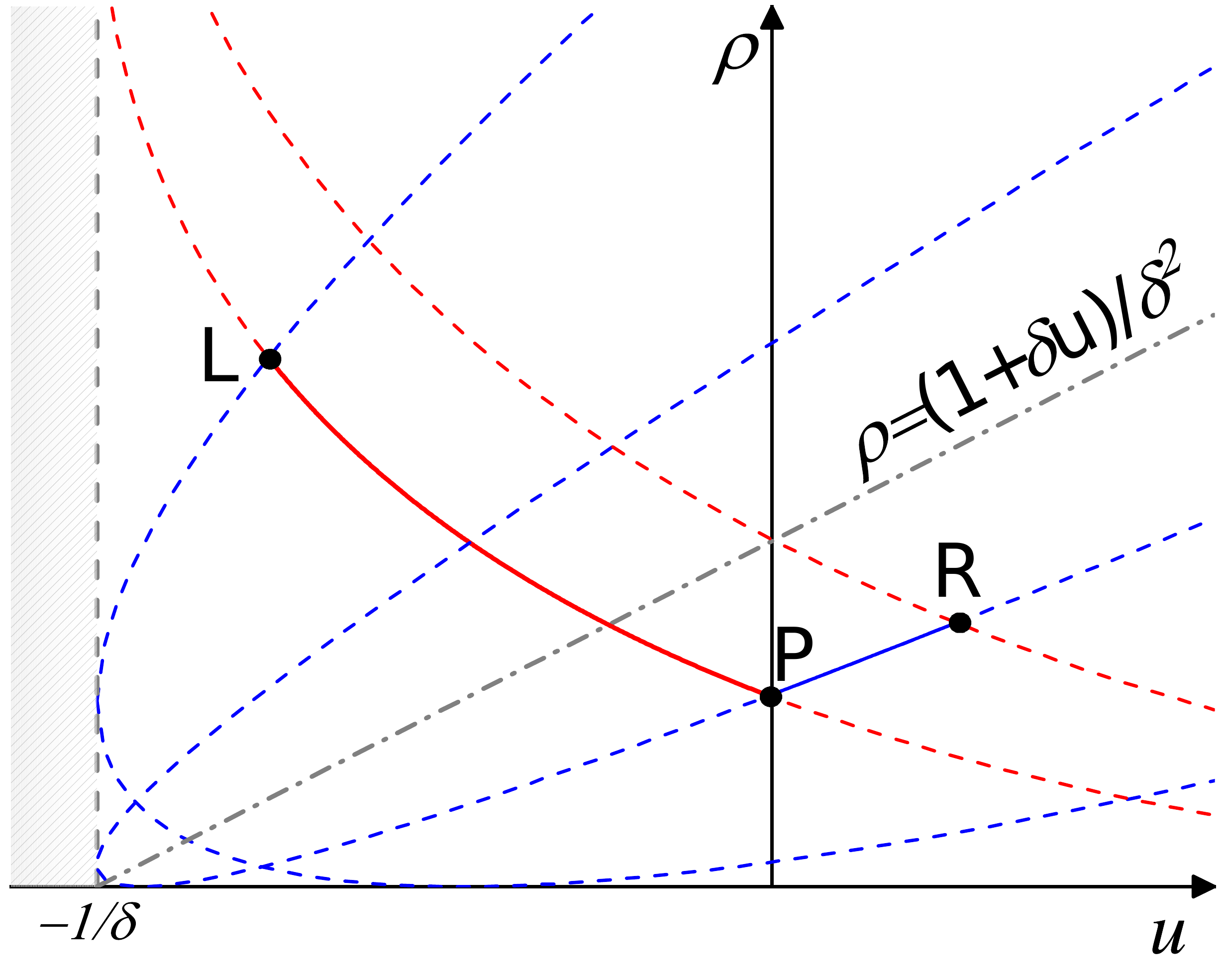}
\includegraphics[width=0.4\textwidth]{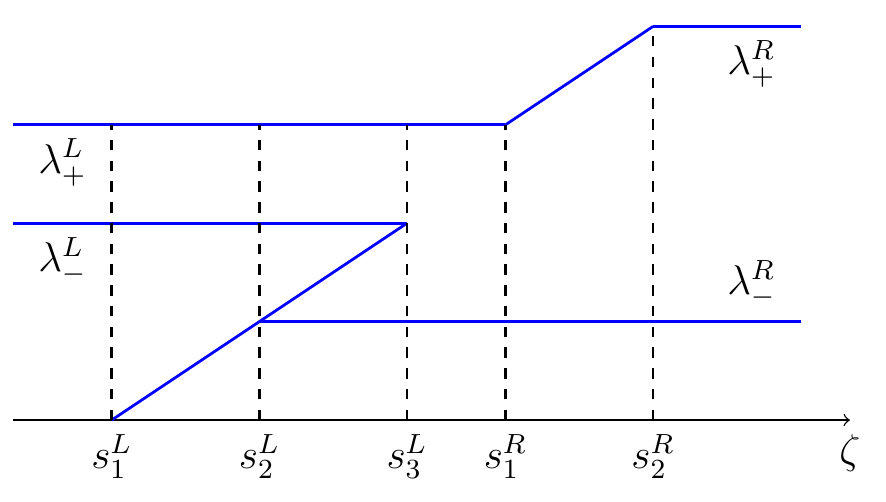}
\caption{(Top panel) Relation between $\rho$ and $u$ for the simple wave solutions in the dispersionless regime for $\rho_L=1.2$, $u_L=-0.8$, $\rho_R=0.6$ and $u_R=0$. (Bottom panel) Diagram representing the evolution of the Riemann invariants as functions
of $\zeta=x/t$ for complex structure.}
\label{Fig16}
\end{figure}

As an example of complex wave structure, we take the boundary conditions of the form $\rho_L=1.2$, $u_L=-0.8$, $\rho_R=0.6$, $u_R=0$ with $\delta=1$. It can be seen from the top panel of Fig.~\ref{Fig16} that such initial conditions lie in different monotonicity regions. This means that one of the waves must consist of a contact DSW or of a combined shock wave. We have a plateau between the waves indicated by a single point $P$ in Fig.~\ref{Fig16}. This plateau is characterized by two relations between Riemann invariants $r_-^P=r_-^R$ and $r_+^P=r_+^L$. Calculating the dispersionless Riemann invariants, we arrive at the diagram shown in the bottom panel of Fig.~\ref{Fig16}. It can be seen from this figure that the wave propagating to the left consists of a cnoidal wave ($s_1^L<\zeta<s_2^L$) and a contact wave ($s_2^L<\zeta<s_3^L$). In this case, the right wave consists of a rarefaction wave only ($s_1^R<\zeta<s_2^R$). Substitution of the dispersion Riemann invariants, which are the solution of the Whitham equations, into the periodic solution gives the wave structure, which is shown in Fig.~\ref{Fig17}. For comparison, a numerical solution is shown in red (thick curve). The vertical dashed lines correspond to the velocities $s_1^L$, $s_2^L$, $s_3^L$, $s_1^R$ and $s_2^R$. As we can see, analytical calculations agrees well with numerics.

\begin{figure}[t] \centering
\includegraphics[width=0.5\textwidth]{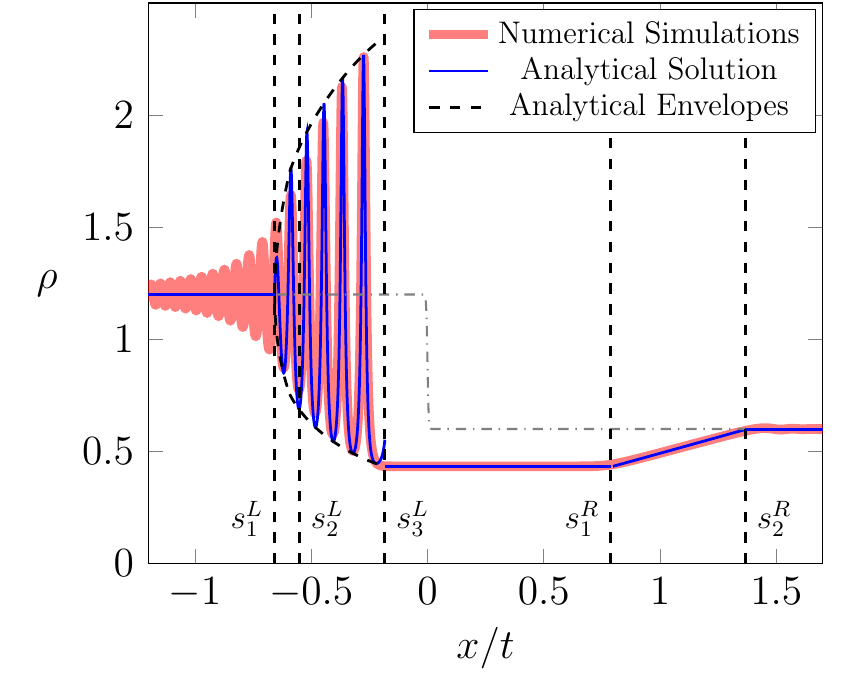}
\caption{Comparison of analytic (blue line) and numerical (red line)
solutions for the initial profile \eqref{init} with $\rho_L=1.2$, $u_L=-0.8$, $\rho_R=0.6$ and $u_R=0$.}
\label{Fig17}
\end{figure}

\begin{figure}[t] \centering
	\includegraphics[width=0.45\textwidth]{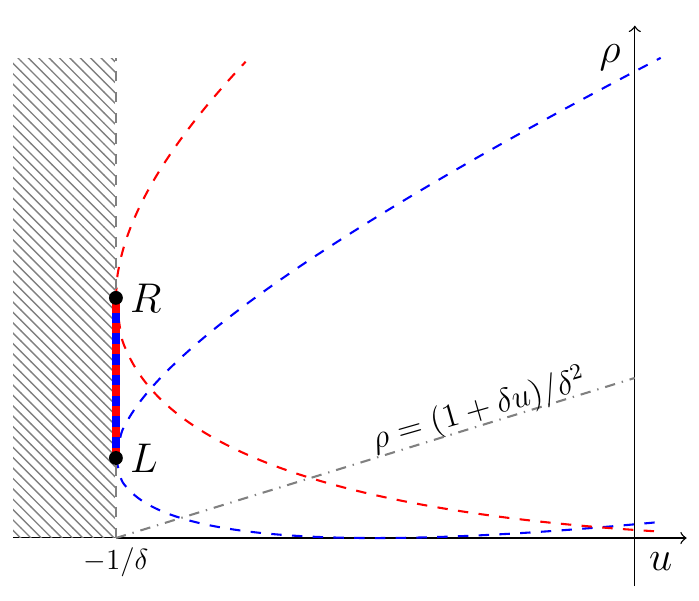}
	\includegraphics[width=0.45\textwidth]{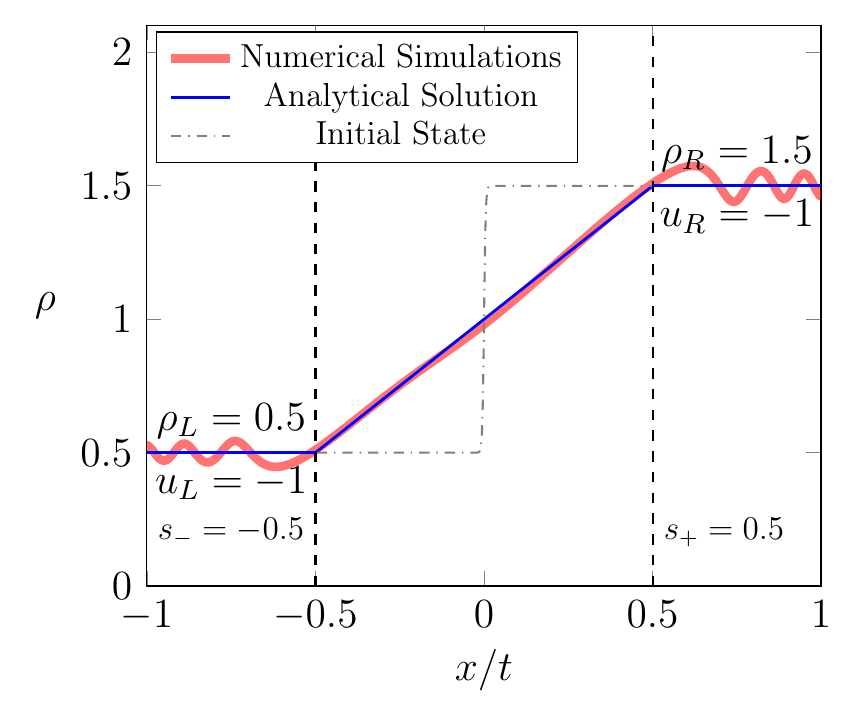}
	\caption{(Top panel) Path in the $(u,I)$-plane for the initial profile \eqref{init} with $\rho_L<\rho_R$ and $u_L=u_R=-1/\delta$. The left and right boundary conditions correspond to points $L$ and $R$ respectively. (Bottom panel) Comparison of analytic (blue line) and numerical (red thick line) solutions for $\rho_L=0.5$, $u_L=-1$, $\rho_R=1.5$ and $u_R=-1$ with $\delta=1$.}
	\label{Fig18}
\end{figure}

	{{We also give an example of a structure that illustrates the distinguishing feature of the generalized CLL equation (\ref{gCLL}). Let us take following boundary conditions: $\rho_L=0.5$, $u_L=-1$, $\rho_R=1.5$ and $u_R=-1$ with $\delta=1$. In this case, the path in the $(u,I)$-plane lies on the border with the region of modulation instability $u<-1/\delta$ (see the top panel in Fig.~\ref{Fig18}). Then the chirp $u$ is constant and is equal to $u=-1/\delta=-1$. The Riemann invariants on the boundaries are equal: $r^L_+=r^L_-$ and $r^R_+=r^R_-$. We can assume that for such initial conditions the Riemann invariants are equal everywhere. Then we get that the dependence of light intensity $\rho$ on $x$ is given by the first equation (\ref{InstPlateau}). A comparison of the numerical calculations with the analytical solution is shown in the bottom panel of Fig.~\ref{Fig18}. One can easily calculate the velocities of the wave edges
			\begin{equation}\label{}
			\begin{split}
			s_-=-\frac{1}{\delta}+\delta\rho_L, \qquad
			s_+=-\frac{1}{\delta}+\delta\rho_R.
			\end{split}
			\end{equation}
	The vertical dashed lines correspond to these velocities.
	}}

\section{Conclusion} \label{sec6}

In this work, the propagation of sufficiently long pulses
in fibers is described by the generalized Chen-Lee-Liu equation, which is related to the class of
the nonlinear Schr\"odinger equations
modified by a self-steepening term.
The Riemann problem of evolution of an initial
discontinuity is solved for this specific case of non-convex dispersive hydrodynamics. It
is found that the set of possible wave structures is much richer than in the convex case
and includes, as structural elements, trigonometric
shock combined with rarefaction waves or cnoidal dispersive shocks.
In the resulting scheme, one solution of the Whitham equations corresponds
to two different wave patterns, and this correspondence is provided by a two-valued
mapping of Riemann invariants to physical modulation parameters. To determine the
pattern evolving from the given discontinuity, we have developed a graphical method.

In principle, one may hope that the results found here
can be observed experimentally in systems similar to that
used in the recent experiment \cite{xckmt-2017}.
However, one should keep in mind that in standard fibers
in addition to the self-steepening effect,
the Raman effect also occurs.
However, the manifestations of these two effects are quite
different and therefore they can be identified separately.
As was shown in the article \cite{IvanovKamchatnov-19},
the main consequence of the Raman
scattering is the formation of stationary shock waves at finite length,
while self-steepening leads to the formation of complex wave structures.
At the same time, the generalized Chen-Lee-Liu theory is also used in the
investigation of modulated wave dynamics of propagation through a single nonlinear
transmission network, which presents some practical interest \cite{Lin-2019}
(see also \cite{KLNV-10} where dispersive shock waves in transmission networks were studied).
The method presented here
is quite flexible and was also applied to other systems with non-convex hydrodynamics \cite{ik-2017,kamch-2018,ikcp-2017}.

\begin{acknowledgments}
I would like to thank my teacher Anatoly Kamchatnov for introducing me to this topic and
for useful discussions during the work on this paper.
The reported study was funded by RFBR, Project number 19-32-90011.
\end{acknowledgments}

\end{document}